\documentclass[twocolumn,10pt]{article}

\usepackage{graphicx}
\usepackage{booktabs}
\usepackage{flushend}
\usepackage{subcaption}
\usepackage{amsmath}
\usepackage{amsthm}
\usepackage{amssymb}
\usepackage{bm}
\usepackage{multirow}
\usepackage{array}
\usepackage{xcolor}
\usepackage{enumitem}
\usepackage{geometry}
\usepackage{color}
\newcommand{\shubham}[1]{\textcolor{black}{#1}}

\begin{document}

\title{Winning an Election: On Emergent Strategic Communication in Multi-Agent Networks}

\author{\textbf{Shubham Gupta} and \textbf{Ambedkar Dukkipati} \\Department of Computer Science and Automation\\Indian Institute of Science, Bangalore - 560012, INDIA.\\\texttt{[shubhamg, ambedkar]@iisc.ac.in}}

\date{}

\maketitle
 
\begin{abstract}
Humans use language to collectively execute abstract strategies besides using it as a referential tool for identifying physical entities. Recently, multiple attempts at replicating the process of emergence of language in artificial agents have been made. While existing approaches study emergent languages as referential tools, in this paper, we study their role in discovering and implementing strategies. We formulate the problem using a voting game where two candidate agents contest in an election with the goal of convincing population members (other agents), that are connected to each other via an underlying network, to vote for them. To achieve this goal, agents are only allowed to exchange messages in the form of sequences of discrete symbols to spread their propaganda. We use neural networks with Gumbel-Softmax relaxation for sampling categorical random variables to parameterize the policies followed by all agents. Using our proposed framework, we provide concrete answers to the following questions: \textbf{(i)} Do the agents learn to communicate in a meaningful way and does the emergent communication play a role in deciding the winner? \textbf{(ii)} Does the system evolve as expected under various reward structures? \textbf{(iii)} How is the emergent language affected by the community structure in the network? To the best of our knowledge, we are the first to explore emergence of communication for discovering and implementing strategies in a setting where agents communicate over a network.\footnote{A shorter version of this paper has been accepted as an extended abstract at AAMAS 2020.}
\end{abstract}

\section{Introduction}
\label{section:introduction}
\begin{quote}
    ``[\dots] only \textit{Sapiens} can talk about entire kinds of entities that they have never seen, touched, or smelled. [\dots] Many animals and human species could previously say `Careful! A lion!' Thanks to the Cognitive Revolution, \textit{Homo sapiens} acquired the ability to say, `The lion is the guardian spirit of our tribe.' This ability to speak about fictions is the most unique feature of Sapiens language [\dots] You could never convince a monkey to give you a banana by promising him limitless bananas after death in monkey heaven.'' \cite{Harari:2015:SapiensABriefHistoryOfHumankind}
\end{quote}

As Yuval Noah Harari points out in \cite{Harari:2015:SapiensABriefHistoryOfHumankind}, language has served a more fundamental purpose in human evolution as opposed to just being used as a referential tool for identifying physical concepts. Among other things, people share ideas, negotiate and devise strategies by using language. In this paper, we study a game played by a set of interconnected agents that communicate using sequences of discrete symbols (an emergent language) to formulate strategies that maximize their rewards.

Recently, in the context of multi-agent reinforcement learning, several attempts at understanding the emergence of language have been made \cite{LazaridouEtAl:2016:MultiAgentCooperationAndTheEmergenceOfNaturalLanguage,MordatchEtAl:2018:EmergenceOfGroundedCompositionalLanguageInMultiAgentPopulations,CaoEtAl:2018:EmergentCommunicationThroughNegotiation}. The general setting is usually viewed as some form of a game where agents are the players. These agents are allowed to communicate with each other using a sequence of discrete symbols (we will refer to these symbols as \textit{words}) that come from a finite set called \textit{vocabulary}. The game is designed in a way so that communication among agents is encouraged to maximize their rewards. Like many existing approaches, we use discrete communication  as it is easier to analyze and is closer to the way humans talk \cite{StuddertKennedy:2005:HowDidLanguageGoDiscrete}.

Most existing approaches use variants of Lewis signaling game \cite{Lewis:1969:LewisSignalingGame}. As an example, in \cite{HavrylovEtAl:2017:EmergenceOfLanguageWithMultiAgentGamesLearningToCommunicateWithSequencesOfSymbols}, there are two agents in the game - sender and receiver. The sender sees an image and transmits a sequence of discrete symbols to describe it. This sequence is decoded by the receiver to understand the intent of sender and select the correct image out of a set of $K$ distinct images. The agents act cooperatively and get a reward each time the receiver is successfully able to pick the correct image. 

In all approaches mentioned above, the language is emergent, i.e. agents have \textit{developed} it from scratch for the sole purpose of maximizing their rewards. The setting of referential games like Lewis signaling game encourages development of a \textit{grounded} language, i.e. words correspond to physical concepts, but, as mentioned earlier, humans also use language for collectively devising strategies, in which case, abstract concepts also play an important role. Recently, a few approaches that study the emergence of language for planning have also been proposed \cite{MordatchEtAl:2018:EmergenceOfGroundedCompositionalLanguageInMultiAgentPopulations,CaoEtAl:2018:EmergentCommunicationThroughNegotiation,BoginEtAl:2018:EmergenceOfCommunicationInAnInteractiveWorldWithConsistentSpeakers} and we briefly describe them in Section \ref{section:related_work}. In this paper, we consider the second setting.

The proposed game involves $n$ agents (which we call members) and two special agents (which we call candidates). Members are connected to each other via an underlying network. At each time step, members broadcast a message in the form of a sequence of discrete symbols to their immediate neighbors and similarly candidates broadcast a message to the members that chose to follow them at the beginning of that time step (more details are given in Section \ref{section:problem_setup}). After $T$ time steps, voting is conducted where each member votes for exactly one candidate. We consider different objectives for agents in this game. For example, one natural objective for each candidate would be to maximize the number of votes that they secure.

At each step, all members update their private preferences for candidates based on the messages that they have received at that time step. These preferences are used for the final voting. Moreover, the messages broadcasted by members are also dependent on their preferences. One can say that, over time, the candidates are supposed to \textit{persuade} members to vote for them using the messages that they broadcast. But since members can also exchange messages with other members, other interesting strategies may also emerge (see Section \ref{section:experiments}).

\shubham{\textbf{Contributions:} \textbf{(i)} We propose a voting game (that can be either competitive or cooperative) to study emergent communication that may or may not be grounded (Section \ref{section:problem_setup}). This is one of our primary contributions in this paper. \textbf{(ii)} We study a setting where communication is restricted along an underlying social network and demonstrate the utility of communication in this setup (Section \ref{section:experiments}).  \textbf{(iii)} We show that the proposed setting can provide interesting insights on emergent strategies, languages and connections between language and network community structure (Section \ref{section:experiments}).}

\textbf{Novelty:} \textbf{(i)} Our proposed voting game is new. \textbf{(ii)} We study emergent languages from the point of view of devising strategies, thus, moving beyond Lewis' referential games. \textbf{(iii)} To the best of our knowledge, we are the first to study emergent languages in a setting where agents are connected via a fixed network topology.

\textbf{Significance:} \textbf{(i)} Communication is essential for achieving artificial general intelligence and study of emergent languages is an important step in this direction. \textbf{(ii)} Network restricted communication is more practical and we highlight the role played by the network topology in our setup. \textbf{(iii)} Our voting game models an interesting real world problem that can encourage future research. \textbf{(iv)} As mentioned above, not all human communication is grounded as humans also use language for abstract strategy development. Thus, studying non-grounded languages is important to understand human communication and build machines that can at some point communicate effectively with humans. Moreover, such communication might also find use in machine-to-machine communication in multi-agent systems.


\section{Voting Game with Communication}
\label{section:problem_setup}

Markov games \cite{Littman:1994:MarkovGamesAsAFormulationForMARL} are commonly used for specification of multi-agent reinforcement learning problems. In this section, we describe the proposed voting game using elements from this framework and highlight the necessary changes that are needed to incorporate inter-agent communication. For a $N$-agent system, a Markov game is specified by the tuple $(\mathcal{S}, \{\mathcal{O}_i, \mathcal{A}_i, r_i\}_{i=1}^N, \mathrm{P}, \gamma)$ where:
\begin{enumerate}[leftmargin=*]
	\item $\mathcal{S}$: Set of all possible states of the environment
	\item $\mathcal{O}_i \,:\, \mathcal{S} \rightarrow \mathcal{Z}_i$: Observation function for $i^{th}$ agent, $\mathcal{Z}_i$ represents its observation space
	\item $\mathcal{A}_i$: Set of actions that can be taken by $i^{th}$ agent
	\item  $r_i \,:\, \mathcal{S} \rightarrow \mathbb{R}$: Reward function for $i^{th}$ agent
	\item $\mathrm{P} \,:\, \mathcal{S} \times \mathcal{A}_1 \times \dots \times \mathcal{A}_N \rightarrow \mathcal{P}(\mathcal{S})$: Transition function. $\mathcal{P}(\mathcal{S})$ denotes the set of all probability distributions over the set $\mathcal{S}$.
	\item $\gamma \in [0, 1]$: Discount factor
\end{enumerate}

The proposed game has two types of agents: $n$ population members $\{M_1, M_2, \dots, M_n\}$ and two candidates $C_1$ and $C_2$, thus $N = n + 2$. We further assume that the members are connected to each other via an underlying social network; the utility of this assumption would soon become clear. A straightforward extension of the proposed framework that accommodates more than two candidates is possible, but we will not discuss it in this paper. 

The two candidates are contesting in an election where they are seeking votes from the members. The game is a finite horizon game consisting of $T$ \textit{propaganda steps} followed by a \textit{voting step}. Each candidate $C_j$ has a fixed \textit{propaganda vector} $\mathbf{c}_j \in \mathbb{R}^d$ and each member $M_i$ has a time dependent \textit{preference vector} $\mathbf{m}_i^{(t)} \in \mathbb{R}^d$. During each propaganda step $t$, each member $M_i$ \textit{follows} one of the two candidates. We use $\mathbf{F}_i^{(t)} \in \{1, 2\}$ to denote the categorical random variable that represents the candidate being followed by member $M_i$ at time $t$. We also use $\mathbf{F}^{(t)} \in \{1, 2\}^n$ to denote the random vector whose $i^{th}$ entry is $\mathbf{F}_i^{(t)}$. The state of the environment at a given time $t$ is encoded by the collection of preference and propaganda vectors described above along with the adjacency matrix $\mathbf{A} \in \{0, 1\}^{n \times n}$ of the underlying social network and the random vector $\mathbf{F}^{(t)}$.

Members $M_i$ only observe their own preference vector $\mathbf{m}_i^{(t)}$ while candidates $C_j$ also observe the network adjacency matrix $\mathbf{A}$ and random vector $\mathbf{F}^{(t)}$ in addition to observing their own propaganda vector $\mathbf{c}_j$. The observation functions $\mathcal{O}_i$, $i = 1, 2, \dots, N$ are defined to distill this information from the state $\mathbf{s}^{(t)} \in \mathcal{S}$.

Agents can take two types of actions in the proposed game: \textit{communication actions} and \textit{modification actions}. At each step, members select one action of each type, thus selecting a joint action, whereas the candidates only select a communication action. The action sets $\mathcal{A}_i$ for $i=1, 2, \dots, N$ are defined accordingly.

\begin{figure}
    \centering
    \includegraphics[width=0.35\textwidth]{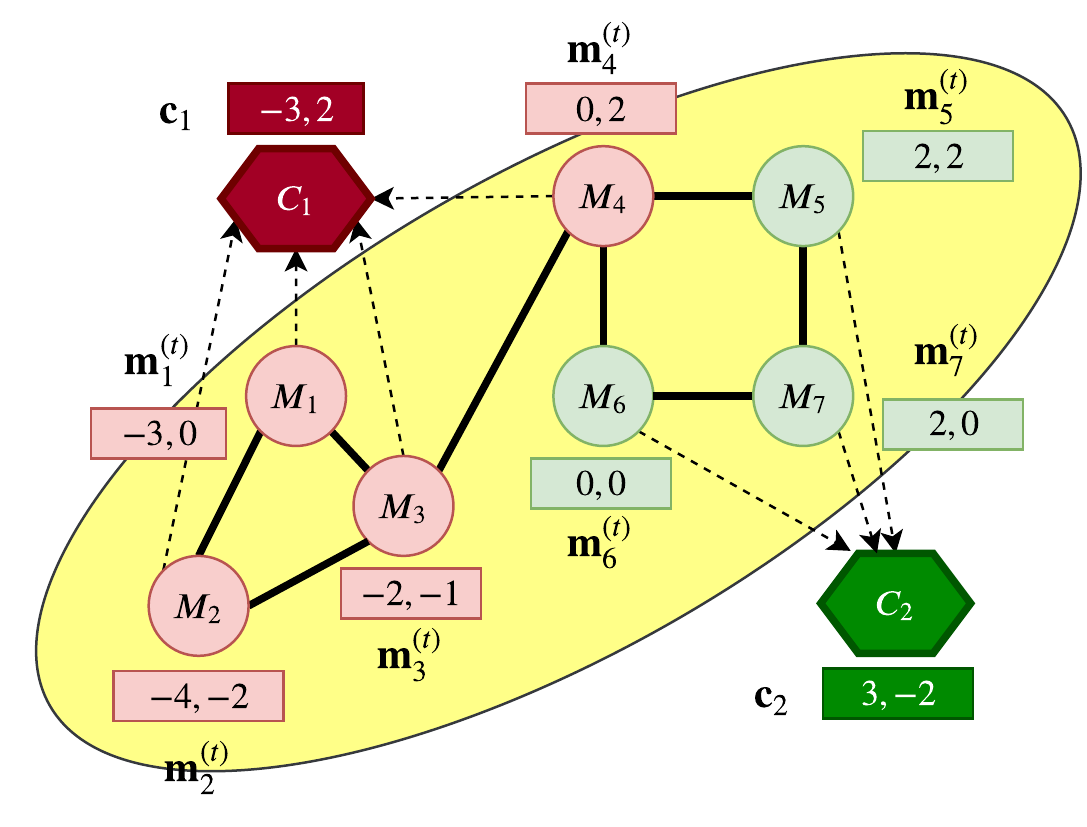}
    \caption{A toy example demonstrating the game. Circles are members and hexagons are candidates. The private \textit{preference}/\textit{propaganda} vectors of members/candidates have also been shown. In this example, these vectors are two dimensional. Members have been colored based on the candidate they follow at the current time step. The yellow ellipse marks the boundary of network, hence candidates are not part of the network. Members broadcast messages to their neighbors, candidates broadcast messages to their followers (represented by dashed arrows). (best viewed in color)}
    \label{fig:game_demo}
\end{figure}

\begin{figure}[t!]
	\centering\includegraphics[scale=0.4]{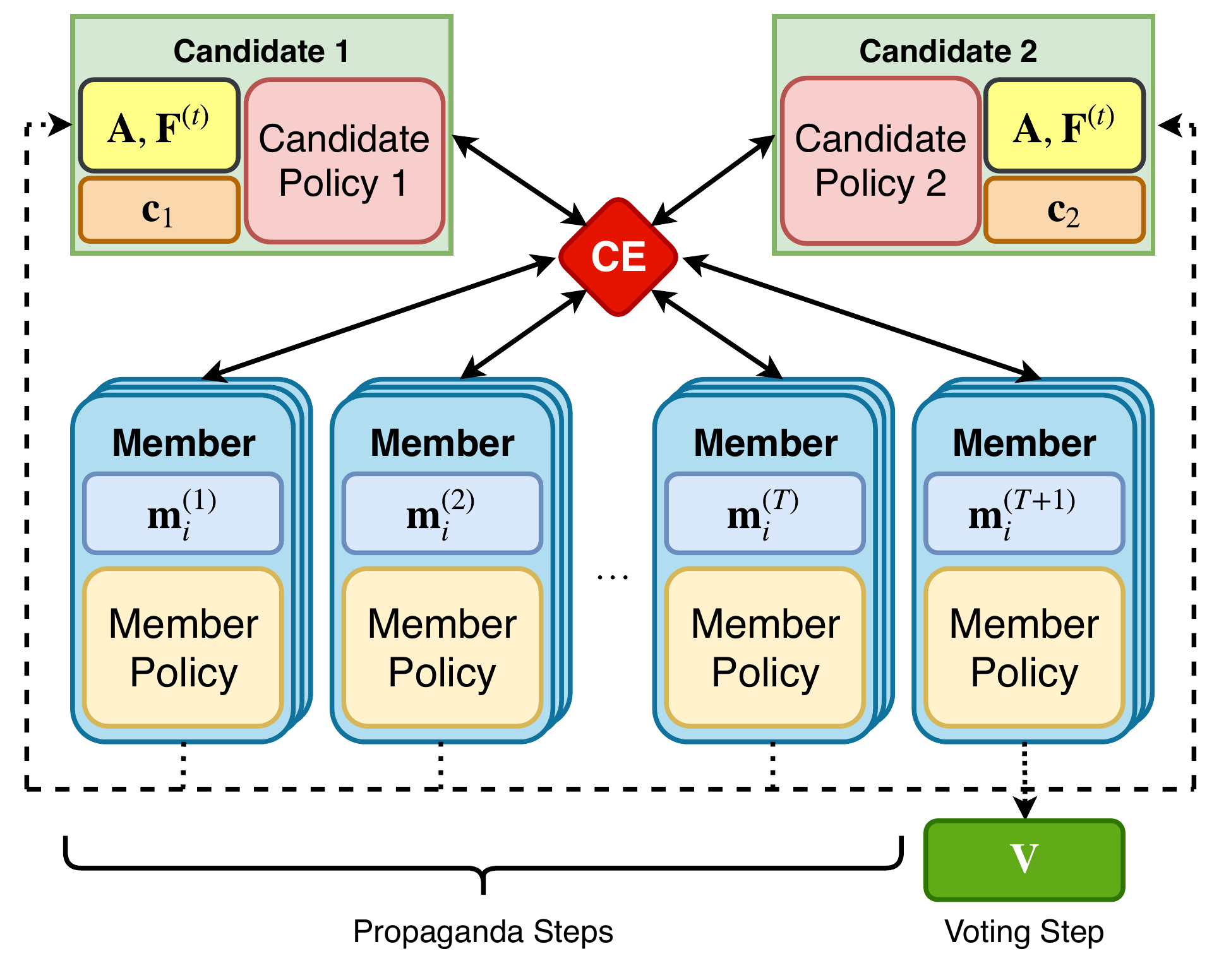}
	\caption{Problem setup: All members share the same policy and have a time dependent preference vector denoted by $\mathbf{m}_i^{(t)}$. Candidates have their own policy and a propaganda vector given by $\mathbf{c}_j$. Candidate's policy uses the network structure $\mathbf{A} \in \{0, 1\}^{n \times n}$ and time dependent following $\mathbf{F}^{(t)} \in \{1, 2\}^n$. All communication happens via the shared discrete communication engine denoted by the red diamond.}
	\label{fig:problem_setup}
\end{figure}

Let $\mathcal{V}$ be a finite set containing $n_{\mathrm{vocab}}$ elements. We call this set \textit{vocabulary} and refer to its elements as \textit{words}. A communication action corresponds to a sequence of discrete symbols (or words) from the set $\mathcal{V}$ of length at most $L_{\mathrm{max}}$. We refer to each such sequence as a \textit{message}. Both candidates and members choose a message at each time $t$ based on their local observations. The message chosen by a member is broadcasted to all of its neighbors in the social network and the message chosen by a candidate $C_j$ is broadcasted to all members $M_i$ for which $\mathbf{F}_i^{(t)} = j$, i.e., to all those members that chose to follow candidate $C_j$ at time $t$. Messages broadcasted at time $t$ become part of the observation made by the receiving agent at time $t + 1$. The communication between agents is emergent as the \textit{policies} for choosing communication actions are learned during the training process. As the members can only communicate along the edges of the underlying network, the network topology plays an important role in determining the nature of emergent communication (Section \ref{section:experiments}). Note that although a member can receive multiple message at a given time step, it broadcasts exactly one message. Similarly, candidates do not receive any messages and broadcast exactly one message at each time step.

Members also take a modification action at each time step in addition to taking a communication action. The modification action of member $M_i$ at time $t$ corresponds to the choice of a vector $\hat{\mathbf{m}}_i^{(t)}$ and scalar $\lambda_i^{(t)}$ that are used for modifying the preference vector $\mathbf{m}_i^{(t)}$ of the member. As before, the \textit{policy} used by members only uses the local information available to them to choose $\hat{\mathbf{m}}_i^{(t)}$ and $\lambda_i^{(t)}$. Note that in the case of members, while choosing both communication actions and a modification actions, the messages received from other agents are also part of the observation made by the members.

The transition function $\mathrm{P}$ specifies the evolution of state of the environment in response to the actions taken by the agents. While the propaganda vectors $\mathbf{c}_1$ and $\mathbf{c}_2$, and adjacency matrix $\mathbf{A}$ remain fixed throughout the episode, the preference vectors get updated based on the modification actions chosen by members as follows:
\begin{equation}
\label{eq:pref_vector_update}
\mathbf{m}_i^{(t + 1)} = (1 - \lambda_i^{(t)}) \mathbf{m}_i^{(t)} + \lambda_i^{(t)} \hat{\mathbf{m}}_i^{(t)}.
\end{equation}
Here, $\lambda_i^{(t)} \in (0, \epsilon)$ and $\epsilon \in (0, 1)$ is a hyperparameter. It is used to enforce the prior that preference vectors do not change very quickly. The only part of the state that changes stochastically is the vector $\mathbf{F}^{(t)}$. The environment randomly samples a value of $\mathbf{F}_i^{(t)}$ by passing the vector $\big(\vert\vert \mathbf{m}_i^{(t)} - \mathbf{c}_1 \vert\vert_2^2 / d, \vert\vert \mathbf{m}_i^{(t)} - \mathbf{c}_2 \vert\vert_2^2 / d \big)$ through Gumbel-Softmax \cite{JangEtAl:2017:CategoricalReparameterizationWithGumbelSoftmax,MaddisonEtAl:2017:TheConcreteDistribution} to get a one-hot encoded vector which represents the choice made by member $M_i$ (also see Appendix \ref{appendix:gumbel_softmax}).

After $T$ propaganda steps, voting is conducted during the voting step. Each member $M_i$ votes for exactly one candidate $C_1$ or $C_2$. We use $\mathbf{V}_i$ to denote the categorical random variable that represents the vote cast by member $M_i$ after $T$ propaganda steps, thus $\mathbf{V}_i \in \{1, 2\}$. As in the case of $\mathbf{F}_i^{(t)}$, the value of $\mathbf{V}_i$ is sampled by passing the vector $\big(\vert\vert \mathbf{m}_i^{(T + 1)} - \mathbf{c}_1 \vert\vert_2^2 / d, \vert\vert \mathbf{m}_i^{(T + 1)} - \mathbf{c}_2 \vert\vert_2^2 / d \big)$ through Gumbel-Softmax to get a one-hot encoded vector. The reward functions $r_i$ for $i = 1, 2, \dots, N$ determine whether the agents will be cooperative or competitive. Agents receive no rewards during the propaganda steps and are rewarded based on the vector $\mathbf{V}$ during the voting step. We study different objectives for candidates under this framework in Section \ref{section:experiments}. For example, the candidates may want to act cooperatively to maximize the votes secured by one of them.

As we have a finite horizon problem, we set the value of discount factor $\gamma$ to $1$.

Candidates follow separate policies but all members share the same policy. However, since the policy used by members is a function of their preference vectors, the members can take different actions and hence the setup is fairly expressive. All members and candidates share the same vocabulary, message encoder and decoder (which we collectively call the \textit{communication engine}) for communicating with each other as explained in Section \ref{section:network_architecture}. Due to this, a common language tends to \textit{emerge} that \textbf{(i)} is consistently understood by all agents, and \textbf{(ii)} allows development and implementation of intelligent strategies for effective propaganda. A toy example has been presented in Figure \ref{fig:game_demo}. Figure \ref{fig:problem_setup} depicts the problem setup.

\begin{figure}[t!]
  \centering
  \includegraphics[scale=0.5]{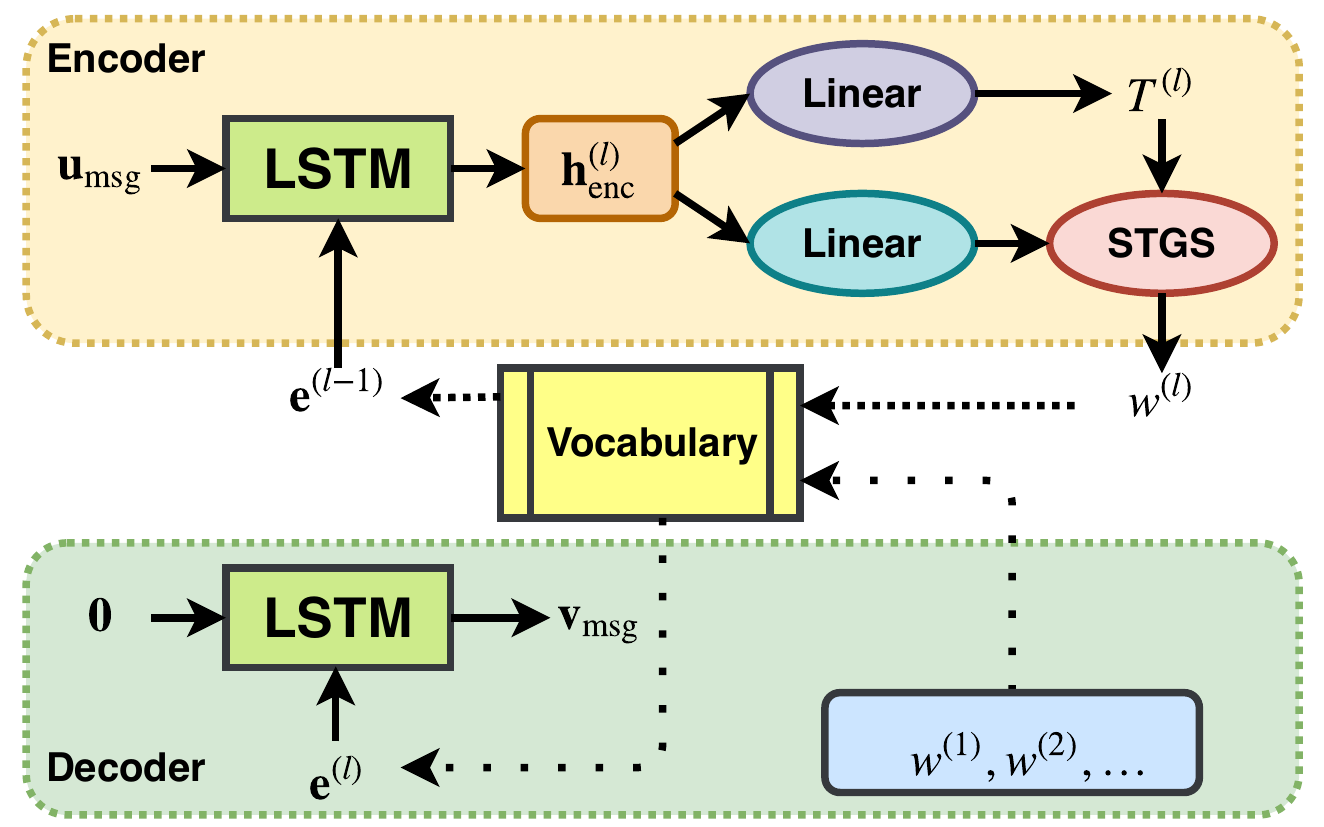}
  \caption{Communication engine: For both LSTMs input is fed from bottom, hidden state is fed from left and output is generated on right side. $w^{(l)}$ is the index $l^{th}$ symbol in the sequence and $\mathbf{e}^{(l)} \in \mathbb{R}^{d_{\mathrm{vocab}}}$ is the corresponding embedding. STGS refers to Straight Through Gumbel-Softmax.}
  \label{fig:communication_engine}
\end{figure}


\section{Policies and Communication Engine}
\label{section:network_architecture}
The members and candidates follow their learned policies for $T$ propaganda steps after which they receive a reward based on the outcome of voting step. The goal of each agent is to maximize their reward. As shown in Figure ~\ref{fig:problem_setup}, our setup consists of four modules: \textbf{(i)} a communication engine that is shared by all agents, \textbf{(ii)} a member policy network that is shared by all members and \textbf{(iii)} one candidate policy network for each of the two candidates. We use the Gumbel-Softmax trick \cite{JangEtAl:2017:CategoricalReparameterizationWithGumbelSoftmax} whenever we need to sample a categorical random variable, this ensures that the setup is end-to-end differentiable.


\subsection{Communication Engine}
\label{section:communication_engine}
The communication engine consists of a shared vocabulary, an encoder and a decoder. The shared vocabulary is a set of $n_{\mathrm{vocab}}$ learnable embeddings each of dimension $d_{\mathrm{vocab}}$, where $n_{\mathrm{vocab}}$ is the number of symbols in the vocabulary. 

The encoder network consists of a LSTM \cite{HochreiterEtAl:1997:LSTM} and two linear layers. It takes a $d_{\mathrm{msg}}$ dimensional vector representation of a message, $\mathbf{u}_{\mathrm{msg}} \in \mathbb{R}^{d_{\mathrm{msg}}}$, as input and produces a sequence of discrete symbols corresponding to it as output. The cell state of LSTM is initialized with $\mathbf{u}_{\mathrm{msg}}$. To get the $l^{th}$ symbol, $w^{(l)}$, in the output sequence, the hidden state $\mathbf{h}_{\mathrm{enc}}^{(l)}$ is passed through a linear layer and a sample from the vocabulary (union a special end token) is obtained using the \textit{Straight Through Gumbel-Softmax} trick \cite{JangEtAl:2017:CategoricalReparameterizationWithGumbelSoftmax,MaddisonEtAl:2017:TheConcreteDistribution}. We use Gumbel-Softmax so that the architecture is end-to-end differentiable which allows the use of standard backpropagation algorithm. 
An alternate approach would be to use policy gradients \cite{Williams:1992:SimpleStatisticalGradientFollowingAlgorithmsForConnectionistReinforcementLearning} but recent similar approaches have demonstrated that Gumbel-Softmax leads to superior performance \cite{HavrylovEtAl:2017:EmergenceOfLanguageWithMultiAgentGamesLearningToCommunicateWithSequencesOfSymbols}. 

The temperature parameter $T_{\mathrm{enc}}^{(l)}$ for Gumbel-Softmax is obtained separately for each symbol in the sequence by passing $\mathbf{h}_{\mathrm{enc}}^{(l)}$ through a hidden layer as:
\begin{equation}
    \label{eq:encoder_temperature}
    T_{\mathrm{enc}}^{(l)} = T_0 + \ln (1 + \exp(\mathrm{Linear}(\mathbf{h}_{\mathrm{enc}}^{(l)}))).
\end{equation}
Here $T_0$ is a hyperparameter. While sampling $l^{th}$ symbol, the LSTM takes embedding corresponding to sampled symbol from previous step, $\mathbf{e}^{(l - 1)} \in \mathbb{R}^{d_{vocab}}$, as input. At the first step, embedding corresponding to a special start token is used as input to the LSTM. The encoder stops when a special end token is encountered or after $L_{\mathrm{max}}$ steps. 

The decoder network takes a sequence of discrete symbols as input and produces a vector representation of message $\mathbf{v}_{\mathrm{msg}} \in \mathbb{R}^{d_{\mathrm{msg}}}$ as output. It consists of a LSTM that takes the embeddings of symbols in input sequence, $\mathbf{e}^{(l)}$, as input. The last hidden state of LSTM is used as $\mathbf{v}_{\mathrm{msg}}$.


\subsection{Candidate Policy}
\label{section:candidate_policy}
At each step both candidates broadcast a message that is received by all members who follow them at that time step. As opposed to members sharing the same policy, candidates have their own policy network. At time step $t$, a candidate policy network takes the underlying network structure $\mathbf{A} \in \{0, 1\}^{n \times n}$ and the current following $\mathbf{F}^{(t)} \in \{1, 2\}^n$ as input. The output is a message encoding $\mathbf{u}_{\mathrm{msg}}$ that when passed through the encoder described in Section \ref{section:communication_engine} generates the sequence to be broadcasted.

Recall that, $\mathbf{F}_i^{(t)}$ is calculated by passing the vector $\big(\vert\vert \mathbf{m}_i^{(t)} - \mathbf{c}_1 \vert\vert_2^2 / d, \vert\vert \mathbf{m}_i^{(t)} - \mathbf{c}_2 \vert\vert_2^2 / d \big)$ through Gumbel-Softmax to get a one-hot encoded vector. This vector is used as a feature vector for member $M_i$. The feature vectors of all members are passed through a three layer Graph Convolution Network (GCN) \cite{KipfEtAl:2017:SemiSupervisedClassificationWithGraphConvolutionalNetworks} that produces a $n \times d_{\mathrm{msg}}$ matrix as output. This matrix is consolidated into a vector of size $d_{\mathrm{msg}}$ using an adaptive max pooling layer. We denote this vector by $\mathbf{s}^{(t)}$.

To make long term strategic decisions, the candidate policy network is equipped with a LSTM that receives $\mathbf{s}^{(t)}$ as input at propaganda step $t$. The hidden state of LSTM concatenated with the propaganda vector of candidate is passed through two linear layers to generate $\mathbf{u}_{\mathrm{msg}}$, the encoding of message to be broadcast at time $t$. We use ELU activation function \cite{ClevertEtAl:2016:FastAndAccurateDeepNetworkLearningByExponentialLinearUnitsELUs} at all layers in GCN and also after the penultimate linear layer before $\mathbf{u}_{\mathrm{msg}}$ is generated.

The generated message encoding $\mathbf{u}_{\mathrm{msg}}$ is passed to the encoder module of communication engine which generates a sequence of discrete symbols which is then broadcasted.


\subsection{Member Policy}
\label{section:member_policy}
As mentioned earlier, all members share the same policy network. At time step $t$, for each member $M_i$, this network takes preference vector $\mathbf{m}_i^{(t)}$ and messages that $M_i$ has received at time step $t$ as input and produces three outputs: \textbf{(i)} an encoding of message $\mathbf{u}_{\mathrm{msg}} \in \mathbb{R}^{d_{\mathrm{msg}}}$ that $M_i$ will broadcast via the communication engine at time $t$, \textbf{(ii)} a scalar $\lambda_i^{(t)} \in (0, \epsilon)$ and \textbf{(iii)} a vector $\hat{\mathbf{m}}_i^{(t)} \in \mathbb{R}^d$ that will be used to modify the preference vector for $M_i$ using \eqref{eq:pref_vector_update}.

The first task is to convert the received messages to their respective vector encodings by passing them through the decoder module of communication engine. Each encoding is concatenated with $\mathbf{m}_i^{(t)}$ to provide the context in which it must be interpreted. The concatenated vectors are passed through two linear layer with ELU activation and adaptive max-pooling is applied on the result to get a single vector $\bar{\mathbf{v}}_{\mathrm{msg}} \in \mathbb{R}^{d_{\mathrm{msg}}}$ that summarizes the received information.

A LSTM is used in the member policy network to keep track of history. For member $M_i$, the cell state of this LSTM is initialized with $\mathbf{m}_i^{(1)}$ concatenated with a vector of zeros. At each propaganda step it takes the summarized information, $\bar{\mathbf{v}}_{\mathrm{msg}}$, as input. The output of LSTM, $\mathbf{h}_i^{(t)}$ is used to generate the three output quantities described earlier.

The encoding of output message, $\mathbf{u}_{\mathrm{msg}}$, is generated by passing $\mathbf{h}_i^{(t)}$ through two linear layers, the first of which uses ELU activation function. The scalar $\lambda_i^{(t)}$ is obtained by passing $\mathbf{h}_i^{(t)}$ though a single linear layer that produces a sigmoid output. We multiply this sigmoid output by $\epsilon$ to restrict its range to $(0, \epsilon)$. The vector $\hat{\mathbf{m}}_i^{(t)}$ is generated by passing $\mathbf{h}_i^{(t)}$ through two linear layers, again, the first of these two linear layers uses ELU activation function.

Once these outputs have been generated, the preference vector of $M_i$ is updated using \eqref{eq:pref_vector_update}. $M_i$ broadcasts the message obtained by passing the generated $\mathbf{u}_\mathrm{msg}$ to the encoder. This message will be received at time step $t + 1$ by all immediate neighbors of $M_i$.

We use the convention that message embeddings that are fed as input to the encoder module of communication engine are denoted by $\mathbf{u}_{\mathrm{msg}}$ and the message embeddings produced as output by decoder are denoted by $\mathbf{v}_{\mathrm{msg}}$. These embeddings depend on the sender as well as on time but we do not make it explicit in the notation to avoid clutter. For example, the output message encoding obtained from member policy network at time step $t$ for $M_i$ should be denoted by $\mathbf{u}_{\mathrm{msg}, i}^{(t)}$ as this will be fed to the encoder. However, we just use $\mathbf{u}_{\mathrm{msg}}$ to denote it. The intended meaning should be clear from the context.


\begin{figure}
  \centering
  \includegraphics[scale=0.6]{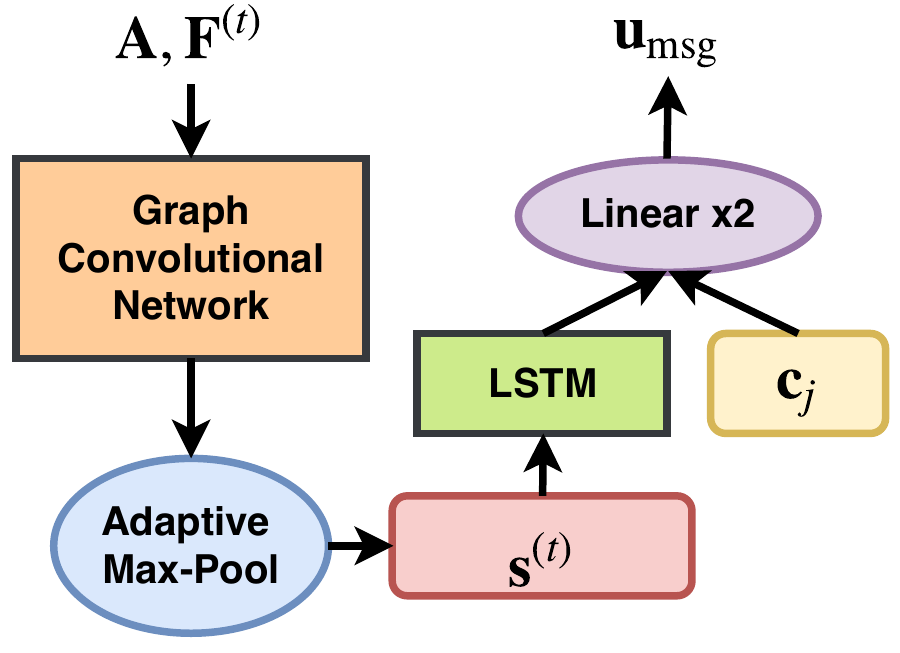}
  \captionof{figure}{Candidate policy network: Graph convolution network computes the current state $\mathbf{s}^{(t)}$ from the observed network and current following. This state is fed to an LSTM that decides the long term strategy to generate the message encoding $\mathbf{u}_{\mathrm{msg}}$.}
  \label{fig:candidate_policy}
\end{figure}

\begin{figure}
\centering
  \includegraphics[scale=0.6]{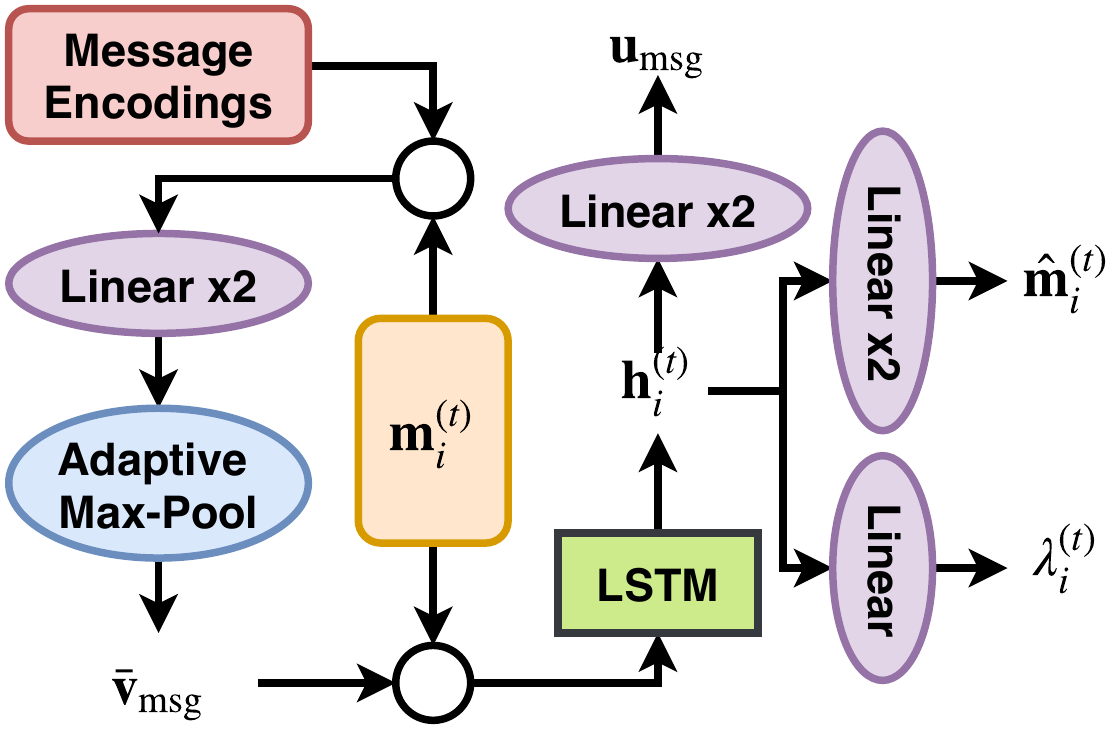}
  \caption{Member policy network: Message encodings from decoder are combined with preference vector $\mathbf{m}_i^{(t)}$ to obtain the input summary $\bar{\mathbf{v}}_{\mathrm{msg}}$ at the bottom left. This is used to generate the three output quantities described in Section \ref{section:member_policy}. A hollow white circle represents concatenation operation.}
  \label{fig:member_policy}
\end{figure}

\subsection{Training Strategy}
\label{section:training_objective}
We simulate \textit{training episodes}, each consisting of $T$ propaganda steps followed by a voting step. We use a constant temperature value $T_{\mathrm{gumbel}}$ for the Gumbel-Softmax distribution used in computing $\mathbf{F}^{(t)}$ and $\mathbf{V}$ as opposed to learning it using an equation similar to \eqref{eq:encoder_temperature}. We experimented with learnable $T_{\mathrm{gumbel}}$ but no significant change was observed. Rewards for members and candidates are computed based on the specific objective function in use as described in Section \ref{section:experiments}. At the end of each training episode we randomly choose to update the policy network of $C_1$, $C_2$ or members with equal probability. Communication engine is updated simultaneously with each policy network update using gradients that are based on the choice of policy network. The setup is end-to-end differentiable due to the use of Gubmel-Softmax and hence backpropagation algorithm can be directly used.



\section{Related Work}
\label{section:related_work}
In the reinforcement learning setting, several attempts at studying problems involving multiple agents that collectively try to solve a common task have been made \cite{Tan:1993:MultiAgentReinforcementLearningIndependentVsCooperativeAgents,BusoniuEtAl:2010:MultiAgentReinforcementLearningAnOverview,SilvaEtAl:2018:AutonomouslyReusingKnowledgeInMultiagentReinforcementLearning}. These agents usually achieve this by sharing information about environment, policies and/or training episodes etc. using a fixed communication protocol. However, recently, multiple approaches that utilize emergent language as a way of communication among agents as opposed to using a fixed ``hard-coded'' communication protocol have been proposed. We briefly discuss a few such approaches here.

Communication can be achieved by exchanging learnable real valued vectors as in \cite{SukhbaatarEtAl:2016:LearningMultiagentCommunicationWithBackpropagation}. One advantage here is that end-to-end differentiability is retained, thus a communication protocol can be \textit{learned} by using standard backpropagation algorithm. Another line of work tries to achieve communication between agents by using sequences of discrete symbols as opposed to real valued vectors with the hope of studying the origin of human language and making the emergent language more ``human like'' \cite{StuddertKennedy:2005:HowDidLanguageGoDiscrete}. Since we use discrete communication, we focus only on approaches of second type.

In \cite{FoersterEtAl:2016:LearningToCommunicateWithDeepMultiAgentReinforcementLearning} differentiable communication is used while training and single bit messages are sent while testing. In \cite{LazaridouEtAl:2016:MultiAgentCooperationAndTheEmergenceOfNaturalLanguage,DasEtAl:2017:LearningCooperativeVisualDialogAgentsWithDeepReinforcementLearning,LazaridouEtAl:2018:EmergenceOfLinguisticCommunicationFromReferentialGamesWithSymbolicAndPixelInput}, variants of Lewis's Signaling Game \cite{Lewis:1969:LewisSignalingGame} have been used - the goal is to use language to describe a target object. As these approaches use language as a referential tool, they employ various strategies to encourage the emergence of a language that is grounded (words represent physical concepts) and compositional (elementary blocks combine to form complex ideas).

In \cite{GauthierEtAl:2016:AParadigmForSituatedAndGoalDrivenLanguageLearning}, the authors argue that an agent can be considered to have learned a language if it can use it to accomplish certain goals. This functional view of language has motivated approaches like \cite{MordatchEtAl:2018:EmergenceOfGroundedCompositionalLanguageInMultiAgentPopulations,CaoEtAl:2018:EmergentCommunicationThroughNegotiation,BoginEtAl:2018:EmergenceOfCommunicationInAnInteractiveWorldWithConsistentSpeakers}. In \cite{MordatchEtAl:2018:EmergenceOfGroundedCompositionalLanguageInMultiAgentPopulations}, each agent partially observes the state and agents have to communicate in a cooperative manner to accomplish tasks like ``move agent 1 to red square''. In \cite{CaoEtAl:2018:EmergentCommunicationThroughNegotiation}, \textit{cheap-talk} \cite{CrawfordEtAl:1982:StrategicInformationTransmission,FarrellEtAl:1996:CheapTalk} is used to allow social agents to communicate using an emergent language while negotiating division of goods in a cooperative game theoretic setting. Our approach is also based on a functional view of language. 

More recently, in \cite{DasEtAl:2019:TarMACTargetedMultiAgentCommunication}, targeted communication was studied where agents can choose who to communicate with at each time step. Our work is different from \cite{DasEtAl:2019:TarMACTargetedMultiAgentCommunication} in two ways: \textbf{(i)} we use discrete communication as opposed to \cite{DasEtAl:2019:TarMACTargetedMultiAgentCommunication} where continuous communication is used; and \textbf{(ii)} agents in our setup can only communicate via an underlying network, i.e., certain agents may never directly communicate with each other; thus the agents must learn to perform well in a constrained environment.

In another closely related work \cite{ZhangEtAl:2018:FullyDecentralizedMultiAgentReinforcementLearningWithNetworkedAgents}, decentralized training of agents that reside on a network has been considered. However, this approach uses a fixed communication protocol between agents (sharing value function estimate with neighbors) as opposed to using emergent communication that is the main object of study in this paper.

To the best of our knowledge this is the first work that studies the setting where agents can only communicate via an underlying network using an emergent language. This leads to some interesting observations (Section \ref{section:experiments}). Moreover, \shubham{our voting game setup is novel and} we consider both cooperative and competitive settings as compared to existing approaches that mainly focus on cooperative agents. 


\section{Experiments}
\label{section:experiments}
We explored different options for: \textbf{(i)} candidate rewards, and \textbf{(ii)} structure of underlying network that connects the members. In this section, we describe these options and present our observations.

\textbf{Rewards for members and candidates:} We experimented with two different rewards for candidates. Recall that $\mathbf{V} \in \{1, 2\}^n$ denotes the voting outcome and let $N_j$ be the number of votes obtained by $C_j$. In the first case, for candidate $C_j$, the reward is given by $N_j$. This makes the candidates competitive, i.e. each one of them wants to maximize the number of votes that they get. In the second case, we have cooperative candidates where for $C_1$ the reward is $N_1$ but for $C_2$ the reward is $-N_2$. In this case, both $C_1$ and $C_2$ want to maximize $C_1$'s number of votes. We call the settings with cooperative candidates and competitive candidates as \textit{biased} and \textit{unbiased} settings respectively.

We use the term \textit{follower reward} for the reward given to members. It is computed for member $i$ as:
\begin{equation}
    \label{eq:follower_reward}
    r_{\mathrm{follower}}(i) = -\vert\vert\mathbf{m}_i^{(T + 1)} - c(i)\vert\vert_2^2,
\end{equation}
where $c(i) = \mathbf{c}_j$ for $j$ such that $\mathbf{V}_i = j$. The reward given to member policy network is the average of rewards obtained by members. The name of reward is justified as it encourages loyalty of followers: even if a particular candidate loses the election, nodes that voted for this candidate still get a higher reward by being ``close'' to it (in terms of the Euclidean distance between their preference vector and candidate's propaganda vector). 

\textbf{Network structure:} We consider two setups - one where the underlying network that connects the members is fixed across all training episodes and other where a randomly sampled network is used for each training episode. Note that in either case, within a training episode, i.e. during propaganda steps and voting step, the network remains fixed.

For the first case, we use the largest connected component of a real world network called Network Science Collaborations network \cite{Newman:2006:FindingCommunityStructureInNetworksUsingTheEigenvectorsOfMatrices} that has $379$ nodes and $914$ edges. For the second case we use a variant of random geometric graph model \cite{Penrose:2003:RandomGeometricGraphs} to sample random networks at the beginning of each episode \shubham{(also see Appendix \ref{appendix:geometric_random_graph_sampling})}. All sampled networks have $100$ nodes. In this model, embeddings $\mathbf{e}_i \in \mathbf{R}^d$ are sampled independently for all nodes $i$ from a zero mean multivariate normal distribution with covariance matrix $\mathbf{I_{d \times d}} / d$. Then, an edge is introduced between all pairs $(i, j)$ for which $\vert\vert\mathbf{e}_i - \mathbf{e}_j\vert\vert_2^2 \leq \delta$ for a fixed constant $\delta > 0$. The choice of $\delta$ determines network sparsity as described in Appendix \ref{appendix:geometric_random_graph_sampling}.

When random geometric graph model is used, the initial preference vector of members are set equal to the corresponding embeddings from graph, i.e. $\mathbf{m}_i^{(1)} = \mathbf{e}_i$. When the network is fixed, we set $\mathbf{m}_i^{(1)}$ equal to the $i^{th}$ row of matrix $\mathbf{Z}_d \in \mathbb{R}^{n \times d}$, where $\mathbf{Z}_d$ is the matrix that contains $d$ leading eigenvectors of $\mathbf{A}$ as its columns (each row is normalized to make it a unit length vector). Irrespective of whether the network is fixed or random, propaganda vector $\mathbf{c}_1$ is sampled randomly from a normal distribution with zero mean and covariance $\mathbf{I_{d \times d}}$, $\mathbf{c}_2$ is then set to $-\mathbf{c}_1$.

When the network is fixed across training episodes, the candidates and members can learn policies that exploit the fixed structure of this network. However, when the network itself is randomly sampled each time, learning network specific strategies is not possible. In such cases we expect the agents to learn more general strategies that can be used on previously unseen networks.

\textbf{Hyperparameters:} Table \ref{table:hyperparams} summarizes the values of different hyperparameters used in all experiments presented here. In all cases we executed the training procedure for $10,000$ episodes. We used Adam optimizer \cite{KingmaBa:2014:AdamAMethodforStochasticOptimization} with default parameters for training all modules.


\begin{table}[t]
    \caption{Hyperparameter values used in experiments.}
    \label{table:hyperparams}
    \begin{center}
    \begin{small}
    \begin{tabular}{lcc}
    \toprule
    \textbf{Description} & \textbf{Symbol} & \textbf{Value} \\
    \midrule
    Constant from equation \eqref{eq:encoder_temperature}   & $T_0$ & $0.2$ \\
    Episode length   & $T$ & $5$ \\
    Learning rate    & - & $0.001$ \\
    Message dim.   & $d_\mathrm{msg}$ & $16$ \\
    Max. sequence length   & $L_\mathrm{max}$ & $5$ \\
    Preference vector dim. & $d$ & $2$ \\
    Symbol embedding dim.   & $d_\mathrm{vocab}$ & $16$ \\
    Temperature for computing $\mathbf{V}$ and $\mathbf{F}^{(t)}$   & $T_\mathrm{gumbel}$ & $0.5$ \\
    Vocabulary size   & $n_\mathrm{vocab}$ & $32$ \\
    \bottomrule
    \end{tabular}
    \end{small}
    \end{center}
\end{table}

\begin{table}[t]
    \caption{Each ordered pair represents the fraction of times $C_1$ and $C_2$ won the game respectively in that order. Note that there were ties as well. \textbf{RGG}: Random Geometric Graph, \textbf{NS}: Net-Science Collaborations network}
    \label{table:score_summary}
    \begin{center}
	\begin{small}
	\begin{tabular}{>{\centering\arraybackslash}p{1.2cm}>{\centering\arraybackslash}p{1.2cm}>{\centering\arraybackslash}p{1.6cm}>{\centering\arraybackslash}p{1.6cm}}
		\toprule
		\textbf{Network Used} & \textbf{Active} & \textbf{Biased Training} & \textbf{Unbiased Training} \\
		\midrule
		\multirow{3}{*}{\textbf{RGG}} & $C_1$ & ($\mathbf{0.94}$, $0.04$) & ($\mathbf{0.69}$, $0.25$) \\
		& $C_2$ & ($\mathbf{0.96}$, $0.02$) & ($0.22$, $\mathbf{0.73}$) \\
		& Both & ($\mathbf{0.99}$, $0.01$) & ($0.31$, $\mathbf{0.60}$) \\
		\hline
		\multirow{3}{*}{\textbf{NS}} & $C_1$ & ($\mathbf{0.92}$, $0.08$) & ($\mathbf{0.62}$, $0.38$) \\
		& $C_2$ & ($\mathbf{0.94}$, $0.06$) & ($0.37$, $\mathbf{0.63}$) \\
		& Both & ($\mathbf{0.96}$, $0.04$) & ($0.46$, $\mathbf{0.54}$) \\
		\bottomrule
	\end{tabular}
	\end{small}
    \end{center}
\end{table}

\subsection{Evaluation Procedure}
\label{section:evaluation_procedure}
In order to demonstrate that the agents are learning something meaningful we observe the effect of placing each trained candidate in an environment where only that candidate is active, i.e. the other candidate is not allowed to broadcast messages. Note that all members can still exchange messages irrespective of the candidate they follow. We also perform an analysis of the language generated by candidates and members. To do so, we take the trained agents and record the language generated by them over $100$ test episodes. From this data, we extract statistics like unigram distribution and bigram distribution for both candidates and members. We make qualitative observations about the emergent language based on these statistics.

Additionally, when the underlying network is fixed, we create a $n \times n_{\textrm{vocab}}$ dimensional member-symbol matrix which we denote by $\mathbf{W}$. $\mathbf{W}_{ij}$ counts the number of times $M_i$ uttered the $j^{th}$ symbol across all $100$ test episodes. We convert this matrix to a tf-idf matrix \cite{JurafskyMartin:2009:SpeechAndLanguageProcessing} and then cluster its rows using spectral clustering (using cosine similarity). We use these results for finding patterns in language usage across members in a fixed network. Note that it does not make sense to do the same exercise for random networks due to the absence of node correspondence across training episodes.


\subsection{Observations}
\label{section:observations}

\textbf{On emergent strategies:} Table \ref{table:score_summary} summarizes the outcome of game under different settings. These scores were obtained by aggregating data over $500$ independent test runs. In all experiments, if biased training is used, the bias is in favor of $C_1$. It can be seen that $C_1$ wins over $90\%$ of the games when the training is biased irrespective of the active candidate because both candidates are trying to make $C_1$ win. In the unbiased case, the active candidate wins as expected, albeit with a smaller margin as some nodes may never get to hear a message in favor of the active candidate even after $T$ propaganda steps. We also observed that in some cases when biased training is used $C_1$ learns to stay dormant by not communicating (i.e., broadcasting sequences of length $0$). This is because it starts relying on $C_2$ to push members towards it. In these cases if $C_1$ alone is active, the outcome is usually a tie.

\begin{figure}
	\vskip 0.2in
	\begin{center}
		\centerline{\includegraphics[scale=0.45]{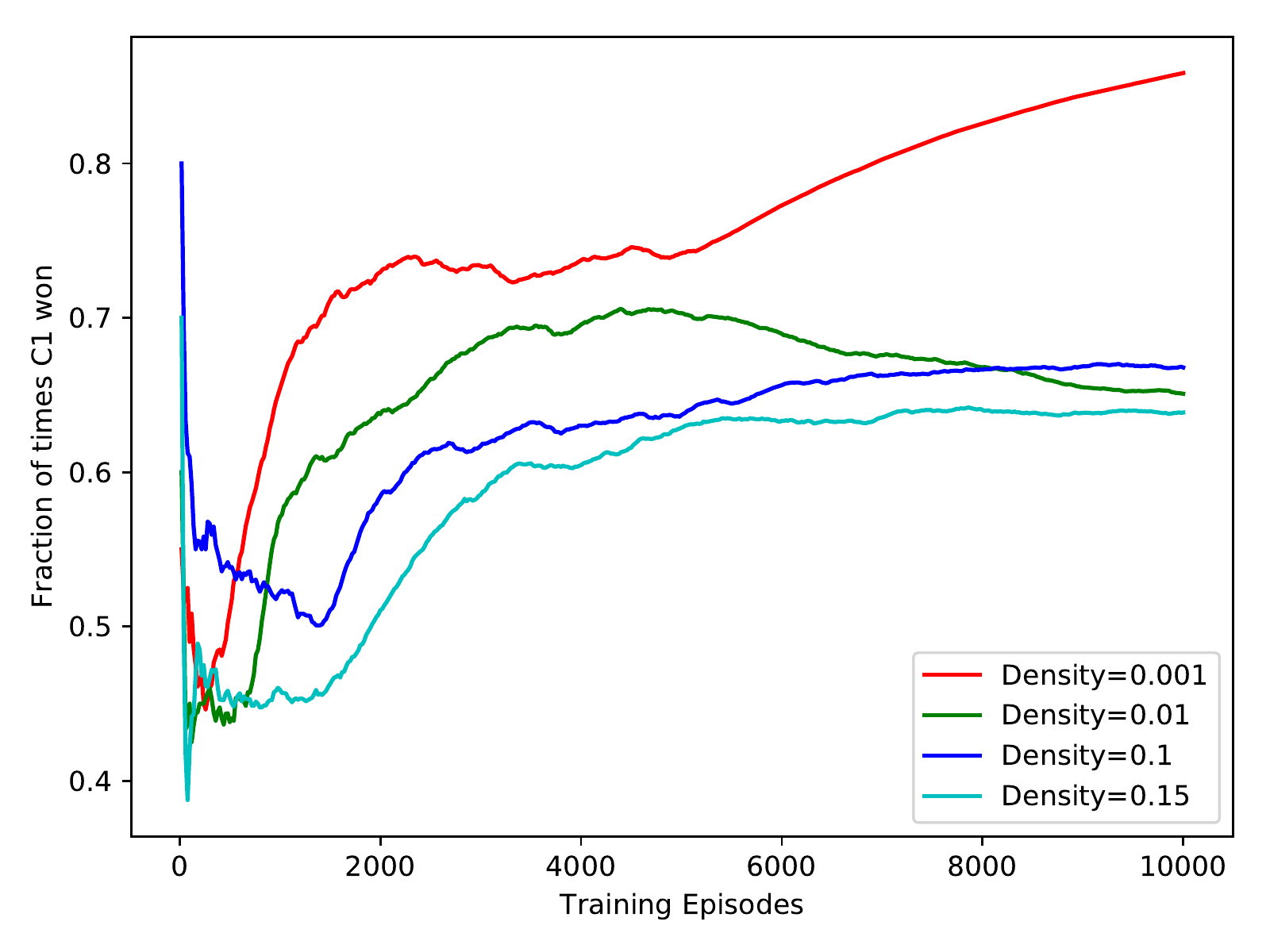}}
		\caption{Fraction of times $C_1$ won the game (biased training) as a function of number of training episodes for different network densities. (best viewed in color)}
		\label{fig:biased_training_progress}
	\end{center}
	\vskip -0.2in
\end{figure}

Fig~\ref{fig:biased_training_progress} shows the training progress for random geometric graphs under biased training. Note that for a given point $t$ on $x$-axis, $y(t)$ represents the fraction of games that $C_1$ has won out of the $t$ games that have been played till that time. Since when the training is in initial stages $C_1$ loses multiple times, the final value at the end of this curve is not same as that reported in Table~\ref{table:score_summary} (we used density = $0.05$ for experiments reported in Table~\ref{table:score_summary}).

\begin{figure*}
	\centering
	\begin{minipage}{0.45\textwidth}
		\centering
		\includegraphics[width=\textwidth]{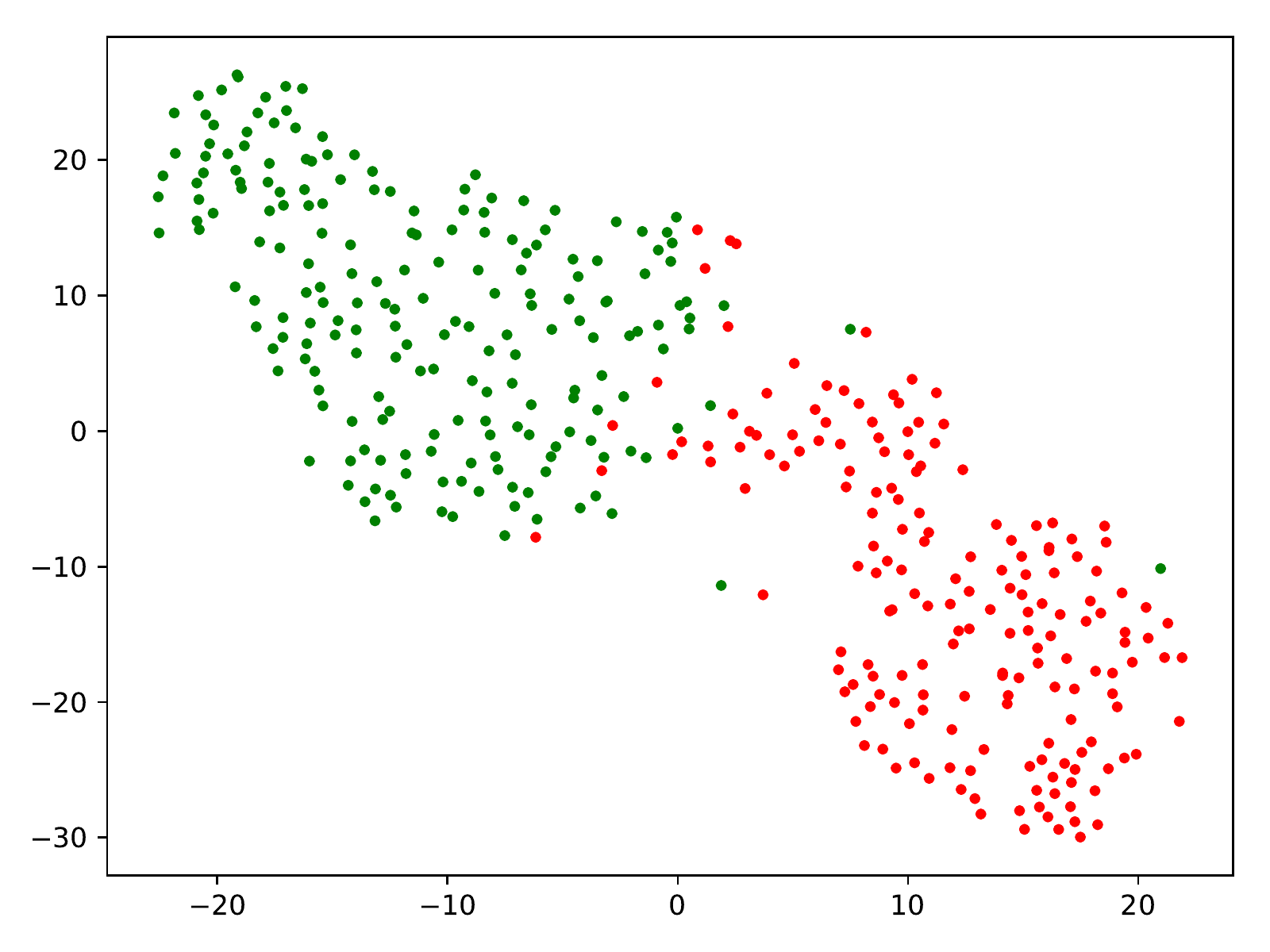}
	\end{minipage}\hfill
	\begin{minipage}{0.35\textwidth}
		\centering
		\includegraphics[width=\textwidth]{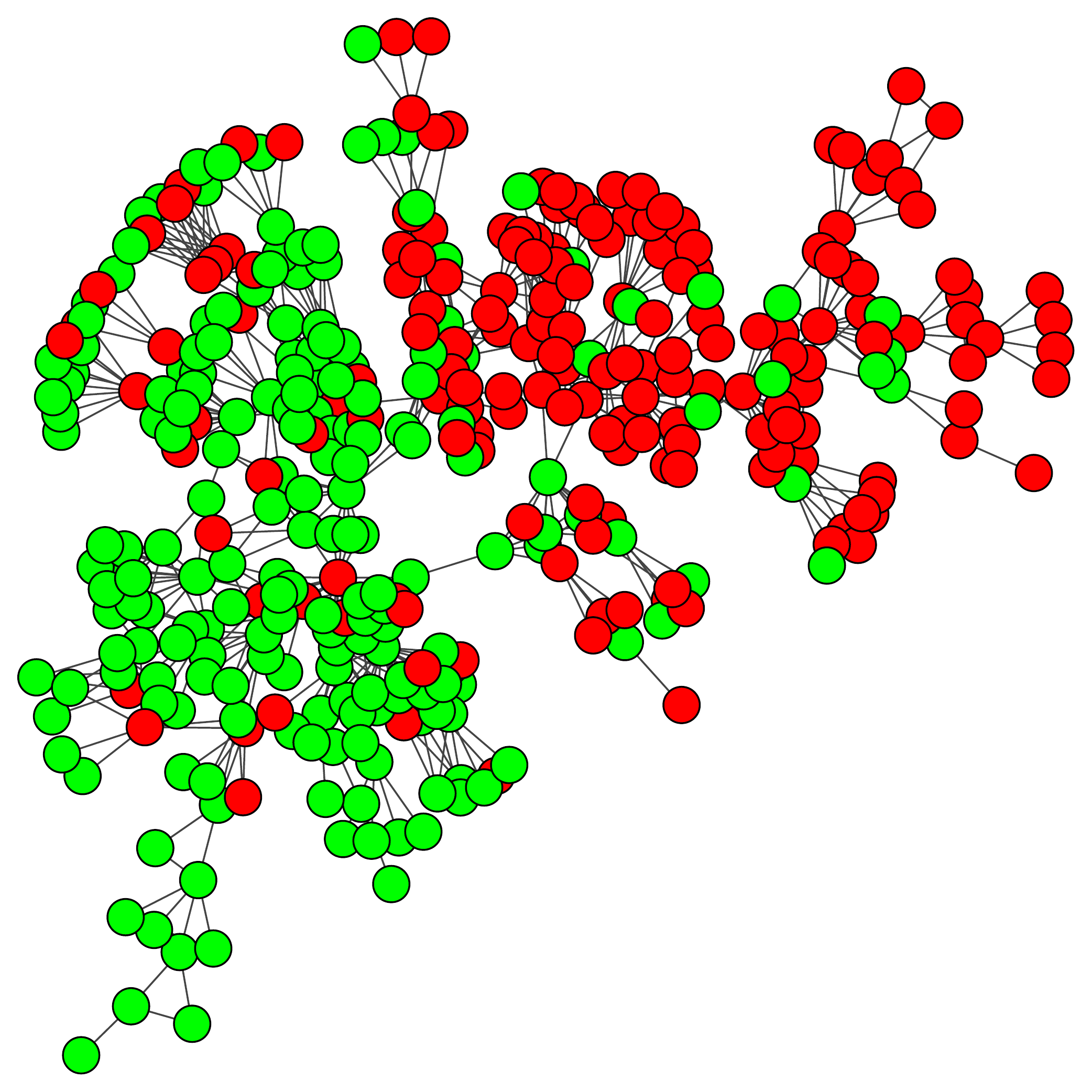}
	\end{minipage}
	\caption{\textbf{Left:} t-SNE plot of rows of language usage matrix $\mathbf{W}$ for Network Science Collaborations network. \textbf{Right: }Clustering of nodes in graph \shubham{\underline{based on their language usage}. Colors represent different clusters. It can be seen that communities based on language usage overlap with structural communities.} (best viewed in color)}
	\label{fig:language_analysis}
\end{figure*} 

Fig~\ref{fig:biased_training_progress} also shows the effect of change in network density \shubham{(Appendix \ref{appendix:geometric_random_graph_sampling})} on model training. When the network is less dense, biased training is easy, since members always receive messages from one of the candidates and there are very few messages being exchanged among members that may distract them. As density increases, communication among members becomes important and since members are rewarded for being loyal (even towards the losing candidate), we believe that it becomes harder to convince them to vote for $C_1$ only. 

\textbf{On emergent languages:} When the training was biased in favor of $C_1$, we observed that preference vectors of members evolved over time to get close to propaganda vector of candidate $C_1$. Additionally, when only one of the candidates was allowed to transmit, say $C_1$, then the following observations were made:
\begin{enumerate}[leftmargin=*]
    \item Members that were following $C_2$ in the first time step did not move during a few initial time steps. This is because the preference vector is updated only if a message is received and when $C_2$ is not allowed to broadcast, these members do not receive any messages until one of their neighbors starts following $C_1$.
    \item Members communicate their intent to their neighbors as evident by the fact that members that were initially following $C_2$ eventually drifted towards $C_1$. Since they did not get a direct message from $C_1$ and $C_2$ was not broadcasting, this can only happen if members communicate meaningfully with each other (also see Appendix \ref{appendix:additional_visualizations}).
\end{enumerate}

We also tried to find patterns in language used by the two candidates. To do this, we looked at the unigram and bigram distributions of the language generated by both candidates. While it is hard to make a quantitative statement, we observed that both candidates use the same high frequency symbols but differ in the usage of low frequency symbols. This is analogues to the high usage of common words like `the', `are', `I' etc. in natural language by everyone and goal specific usage of less common words like `back-propagate', `'inference` etc. by domain experts.

\textbf{On language and network communities:} We also analyzed the language used by all members when the underlying network is fixed as described in Section \ref{section:evaluation_procedure}. Fig~\ref{fig:language_analysis} shows the t-SNE \cite{Maaten:2008:VisualizingDataUsingTSNE} plot of rows of matrix $\mathbf{W}$ and the result of clustering the members (Section \ref{section:evaluation_procedure}) for Network Science Collaborations network based on rows of matrix $\mathbf{W}$. Fig~\ref{fig:language_analysis} was obtained using unbiased training, however similar results are obtained under biased training as well. From the t-SNE plot, one can see a clear difference between language usage in two clusters. Moreover, although these clusters were discovered based on language usage, they naturally correspond to underlying structural communities. This implies that members that are connected to each other develop a language of their own which may be different from language developed in other communities. One can also see some overlap in the t-SNE plot, we hypothesize that these members play the role of translators between the two communities, although further experiments are needed to validate this hypothesis. We observed similar results for another real world network called Polbooks network \cite{Krebs:2004:PoliticalBooksDataset} where nodes correspond to books about US politics. There are $105$ nodes and $441$ edges in this network. Clustering based on language usage yielded communities that exactly overlapped with the two structural communities present in the network. We omit the results here due to space constraints.


\section{Conclusion}
\label{section:conclusion}
In this paper we studied a voting game where the agents learn to communicate with each other through an emergent language with the goal of developing and implementing intelligent strategies that maximize their rewards \shubham{through an effective propaganda}. Further, this communication is allowed over an underlying network that connects the agents. We explored different experimental setups (for example, cooperative vs competitive agents) and presented our observations that, among other things, answer the questions raised at the beginning of this paper:

\textbf{Q1: Do agents learn to communicate in a meaningful way and does the emergent communication play a role in deciding the winner?}  Yes. In Section \ref{section:observations}, we describe how members persuade their neighbors to go to the active candidate when only one candidate is active. This not only shows that the communication is useful, it also shows that candidates have learned to exploit members for spreading their propaganda.

\textbf{Q2: Does the system evolve as expected under various reward structures?} Yes, it is evident from Table \ref{table:score_summary} that agents left in isolation behave as expected in both biased and unbiased training settings.

\textbf{Q3: How is emergent language affected by community structure?} Fig.~\ref{fig:language_analysis} shows that the emergent language in one structural community is different from that in other structural communities in the underlying network.

\shubham{Due to the flexibility of our proposed framework, many interesting questions can be studied under it and we believe that our work will serve as a stepping stone for future research in this direction.} 
For example, one could study the effect of having an underlying network that changes with time as preferences of members evolve (within an episode). Another interesting case would be to have the members compete amongst themselves to secure the highest number of votes as opposed to having designated special candidate agents. Could such a setup explain why communities form in real world networks? Are they a result of globally competing agents with a \shubham{local neighborhood based reward structure}? What if agents are given the ability to privately communicate with each other without broadcasting a message? \shubham{Can other possible reward structures explain different aspects of the corresponding real world problem? Can we discover new insights regarding the connections between different domains of study like network science, game theory and reinforcement learning?}


\bibliographystyle{unsrt}
\bibliography{biblio}


\appendix
\newpage
\textcolor{white}{.}
\newpage


\section{Gumbel-Softmax}
\label{appendix:gumbel_softmax}
Categorical random variables are useful in many situations, however, since the reparameterization trick \cite{KingmaEtAl:2013:AutoEncodingVariationalBayes} can not be applied to them, it is not possible to backpropagate through samples from categorical random variables. Gumbel-Softmax \cite{JangEtAl:2017:CategoricalReparameterizationWithGumbelSoftmax} (also independently discovered by \cite{MaddisonEtAl:2017:TheConcreteDistribution}) offers a continuous relaxation for such categorical distributions which allows one to use it with standard backpropagation algorithm.

Here we will only describe the usage of Gumbel-Softmax as a tool, but we encourage the readers to read the original papers for a more detailed exposition. In the context of this paper, there are three cases where we need to sample from a categorical distribution: \textbf{(i)} While sampling the next symbol in a sequence from the vocabulary, \textbf{(ii)} while choosing a candidate to follow, i.e.
sampling $\mathbf{F}_i^{(t)}$ and \textbf{(iii)} while voting, i.e. sampling $\mathbf{V}_i$. We use the Gumbel-Softmax for all these cases.

Suppose one wishes to sample a categorical random variable $\mathbf{X}$ from a distribution over $K$ elements given by $\bm{\pi} = (\pi_1, \dots, \pi_K)$ where $\pi_k = \mathrm{P}(\mathbf{X} = k)$. To obtain a continuous relaxation, one can instead sample from a $K - 1$ dimensional simplex $\Delta^{K - 1}$ to get a random vector $\mathbf{y} = (y_1, \dots, y_K)$ such that $y_k \geq 0$ and $\sum_k y_k = 1$. The Gumbel-Softmax distribution allows one to sample from $\Delta^{K - 1}$ based on $\bm{\pi}$ as:
\begin{equation}
    \label{eq:gumbel_softmax_sample}
    y_k = \frac{\exp((\log(\pi_k) + g_k) / T_{\mathrm{gumbel}})}{\sum_j \exp((\log(\pi_j) + g_j) / T_{\mathrm{gumbel}})},
\end{equation}
where $T_{\mathrm{gumbel}} > 0$ is the temperature parameter and $g_1, \dots, g_K$ are i.i.d. samples from Gumbel(0, 1) distribution that are obtained as:
\begin{equation}
    \label{eq:gumbel_sample}
    g_k = -\log(-\log(u_k)),
\end{equation}
where $u_1, \dots, u_K$ are i.i.d. samples from Uniform(0, 1).

If the distribution $\mathbf{y} = (y_1, \dots, y_K)$ has most of its mass concentrated at a particular $y_k$, then this vector can be used as an approximation for a one-hot encoded vector that represents the $k^{th}$ discrete element over which the original distribution $\bm{\pi}$ was defined. Loosely speaking, one can show that in the limit of $\mathrm{T_{gumbel}} \rightarrow \infty$, $\mathbf{y}$ becomes a one-hot vector. However as $T_\mathrm{gumbel}$ becomes small, the variance in gradients with respect to $\bm{\pi}$ increases and hence there is a tradeoff.

For a positive value of $\mathrm{T_{gumbel}}$, $\mathbf{y}$ is only an approximation to a one-hot encoded vector but is not actually one-hot encoded. Since we want to be able to communicate via discrete symbols, we would like to use one-hot encoded vectors only. The trick, called \textit{Straight Through Gumbel Softmax}, achieves this by taking the arg-max of $(y_1, \dots. y_K)$ during the forward pass to get an actual one-hot encoded vector, but using the gradients with respect to $(y_1, \dots. y_K)$ as an approximation to the gradients with respect to the one-hot vector during the backward pass.

Let $\hat{\mathbf{y}} = (\hat{y}_1, \dots, \hat{y}_K)$ be a one-hot encoded vector such that $\hat{y}_k = \mathbf{I}\{k = \mathrm{argmax}_j y_j \,\,\,\, \wedge \,\,\,\, k \leq k', \forall k' : k' = \mathrm{argmax}_j y_j\}$, one implements the Straight Through Gumbel-Softmax trick as:
\begin{equation}
    \label{eq:stgs_implementation}
    \mathbf{y}_\mathrm{out} = (\hat{\mathbf{y}} - \mathbf{y})\mathrm{.detach()} + \mathbf{y},
\end{equation}
where $\mathrm{detach()}$ is an operation that prevents the gradients from flowing through the expression on which it was called. Note that since $\mathbf{y}$ was just added and subtracted, $\mathbf{y}_\mathrm{out} = \hat{\mathbf{y}}$, but the gradients that will flow though $\mathbf{y}_\mathrm{out}$ will be equal to the gradients with respect to $\mathbf{y}$.

This allows one to use actual one-hot vectors while still retaining end-to-end differentiability of the model.


\section{Controlling Sparsity of Random Geometric Graph}
\label{appendix:geometric_random_graph_sampling}
Recall that a random geometric graph containing $n$ nodes is sampled as follows: \textbf{(i)} Sample $\mathbf{e}_1, \dots, \mathbf{e}_n \sim \mathcal{N}(0, \mathbf{I}_{d \times d}/d)$ and \textbf{(ii)} for $i < j$, set $\mathbf{A}_{ij} = \mathbf{A}_{ji} = 1$, if $\vert\vert\mathbf{e}_i - \mathbf{e}_j\vert\vert_2^2 \leq \delta$ for a fixed constant $\delta$. We wish to choose $\delta$ such that the network has a desired level of sparsity $\beta = \mathrm{P}(\mathbf{A}_{ij} = 1)$ (note that all entries of matrix $\mathbf{A}$ are identically distributed). We know that:
\begin{align*}
    \beta = \mathrm{P}(\mathbf{A}_{ij} = 1) &= \mathrm{P}(\vert\vert\mathbf{e}_i - \mathbf{e}_j\vert\vert_2^2 \leq \delta) \\
    &= \mathrm{P}(\sum_{k = 1}^d (\mathbf{e}_i - \mathbf{e}_j)_k ^ 2 \leq \delta).
\end{align*}

Since $\mathbf{e}_i, \mathbf{e}_j \sim \mathcal{N}(0, 1/d)$, are independent, $\mathbf{e}_i - \mathbf{e}_j \sim \mathcal{N}(0, 2/d)$ or:
\begin{equation*}
    \sqrt{\frac{d}{2}} (\mathbf{e}_i - \mathbf{e}_j) \sim \mathcal{N}(0, 1).
\end{equation*}

Using the fact that sum of squares of $d$ independent $\mathcal{N}(0, 1)$ random variables is a chi-squared random variable with $d$ degrees of freedom, we get:   
\begin{equation*}
    \frac{2}{d} \sum_{k = 1}^d (\mathbf{e}_i - \mathbf{e}_j)_k ^ 2 \sim \chi^2_d 
\end{equation*}

Thus,
\begin{equation*}
    \mathrm{P}(\sum_{k = 1}^d (\mathbf{e}_i - \mathbf{e}_j)_k ^ 2 \leq \delta) = \mathrm{P}(\mathbf{Z} \leq \frac{d \delta}{2}),
\end{equation*}
where $\mathbf{Z} \sim \chi^2_d$. Let $F_d(.)$ denote the CDF of $\mathbf{Z}$, then we want:
\begin{equation*}
    F_d \Big(\frac{d\delta}{2} \Big) = \beta.
\end{equation*}

Thus, for desired sparsity $\beta$, one can compute $\delta$ as:
\begin{equation*}
    \delta = \frac{2}{d}F_d^{-1}(\beta)
\end{equation*}


\section{Additional Visualizations}
\label{appendix:additional_visualizations}
\begin{figure*}
    \centering
    \begin{subfigure}[t]{0.33\textwidth}
        \centering
        \includegraphics[width=\linewidth]{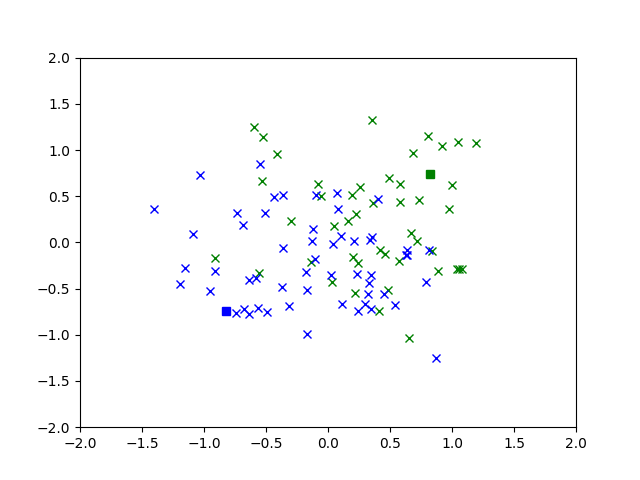}
    \end{subfigure}%
    ~ 
    \begin{subfigure}[t]{0.33\textwidth}
        \centering
        \includegraphics[width=\linewidth]{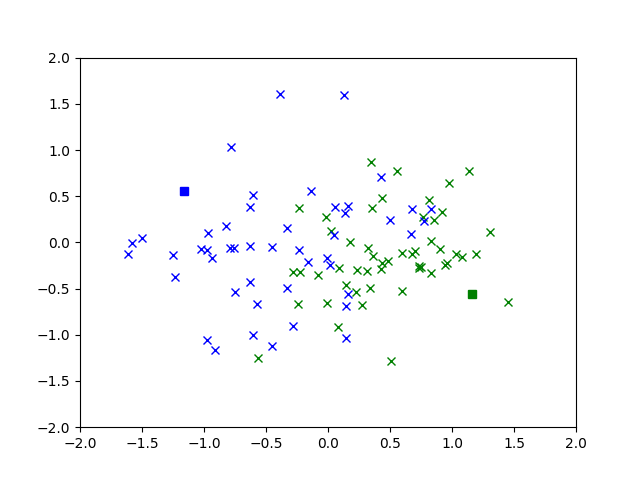}
    \end{subfigure}%
    ~
    \begin{subfigure}[t]{0.33\textwidth}
        \centering
        \includegraphics[width=\linewidth]{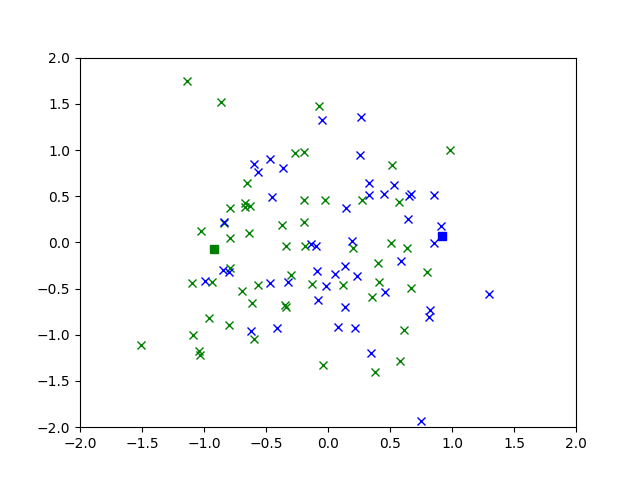}
    \end{subfigure}%
    \\
    \begin{subfigure}[t]{0.33\textwidth}
        \centering
        \includegraphics[width=\linewidth]{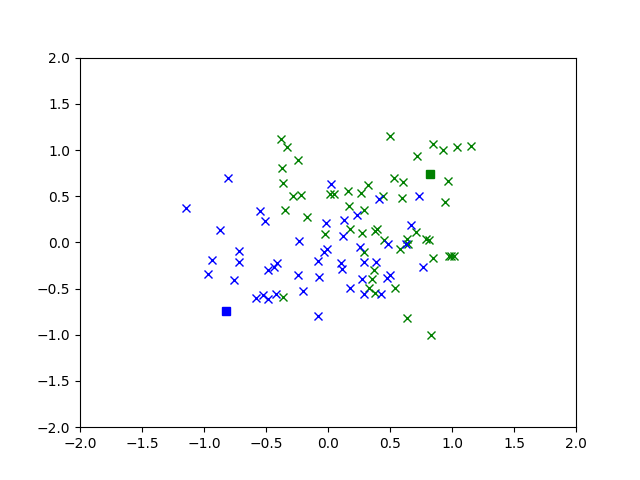}
    \end{subfigure}%
    ~ 
    \begin{subfigure}[t]{0.33\textwidth}
        \centering
        \includegraphics[width=\linewidth]{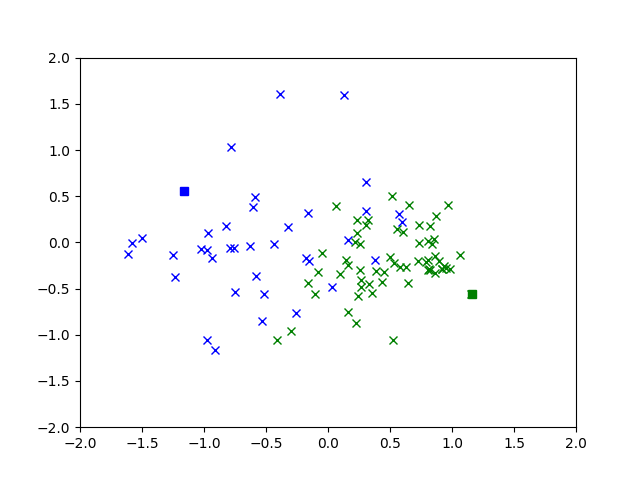}
    \end{subfigure}%
    ~
    \begin{subfigure}[t]{0.33\textwidth}
        \centering
        \includegraphics[width=\linewidth]{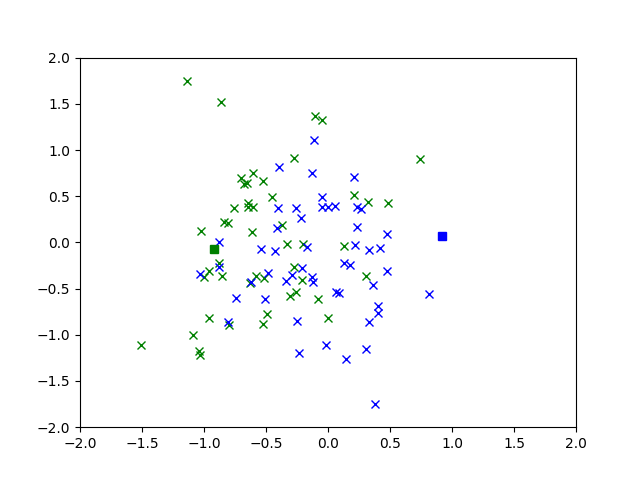}
    \end{subfigure}%
    \\
    \begin{subfigure}[t]{0.33\textwidth}
        \centering
        \includegraphics[width=\linewidth]{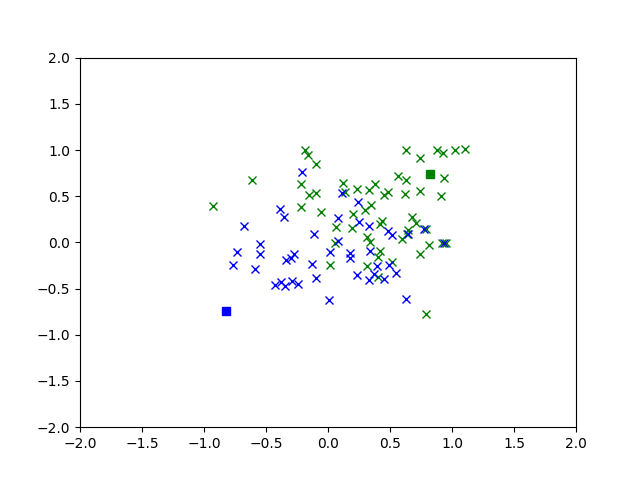}
    \end{subfigure}%
    ~ 
    \begin{subfigure}[t]{0.33\textwidth}
        \centering
        \includegraphics[width=\linewidth]{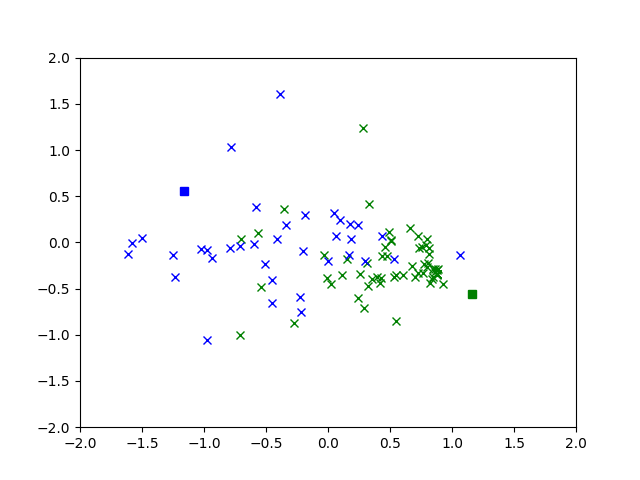}
    \end{subfigure}%
    ~
    \begin{subfigure}[t]{0.33\textwidth}
        \centering
        \includegraphics[width=\linewidth]{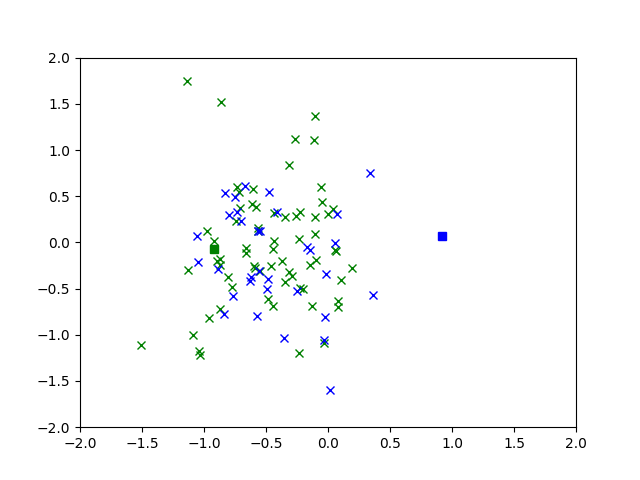}
    \end{subfigure}%
    \\
    \begin{subfigure}[t]{0.33\textwidth}
        \centering
        \includegraphics[width=\linewidth]{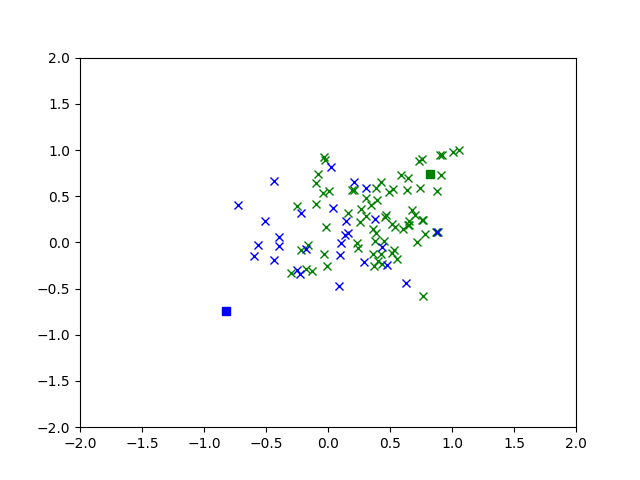}
    \end{subfigure}%
    ~ 
    \begin{subfigure}[t]{0.33\textwidth}
        \centering
        \includegraphics[width=\linewidth]{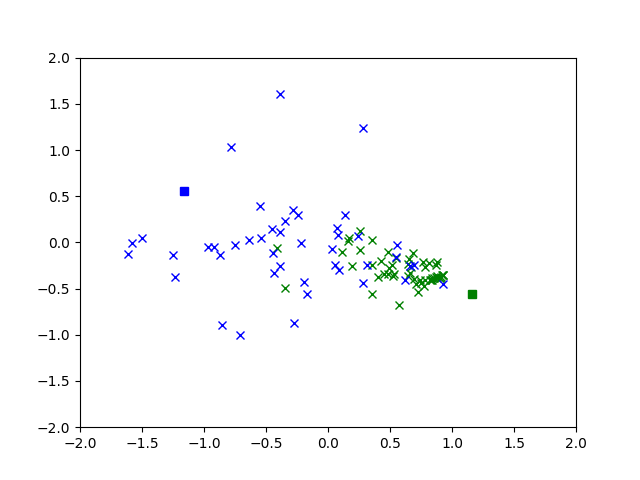}
    \end{subfigure}%
    ~
    \begin{subfigure}[t]{0.33\textwidth}
        \centering
        \includegraphics[width=\linewidth]{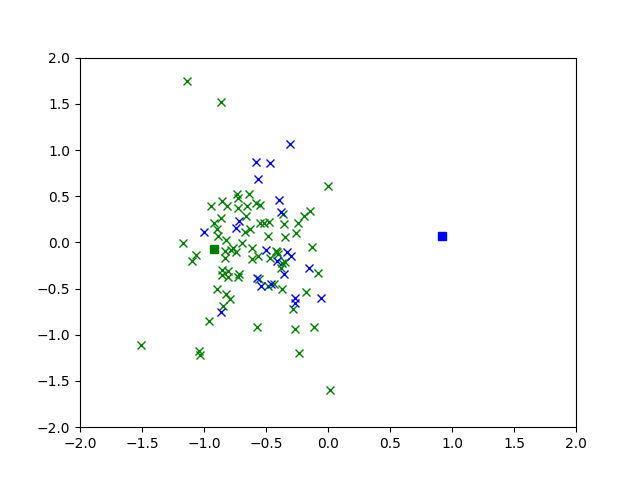}
    \end{subfigure}%
    \\
    \begin{subfigure}[t]{0.33\textwidth}
        \centering
        \includegraphics[width=\linewidth]{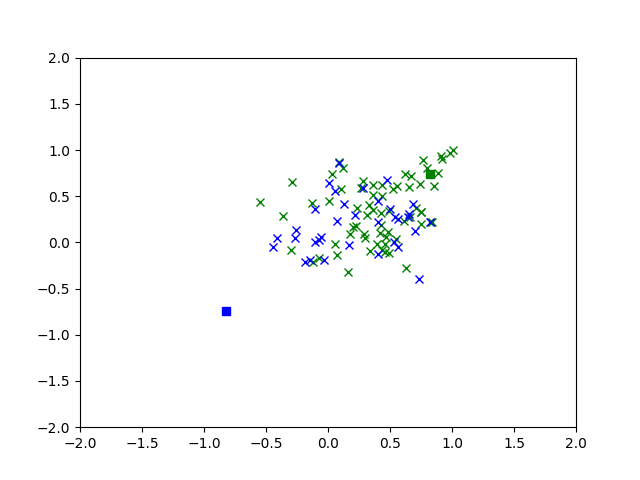}
    \end{subfigure}%
    ~ 
    \begin{subfigure}[t]{0.33\textwidth}
        \centering
        \includegraphics[width=\linewidth]{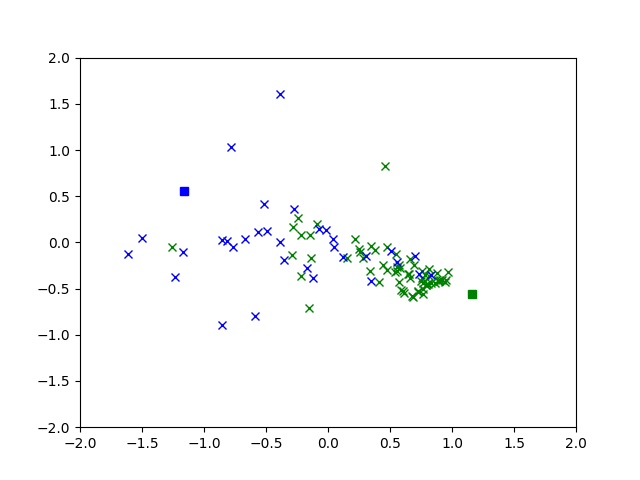}
    \end{subfigure}%
    ~
    \begin{subfigure}[t]{0.33\textwidth}
        \centering
        \includegraphics[width=\linewidth]{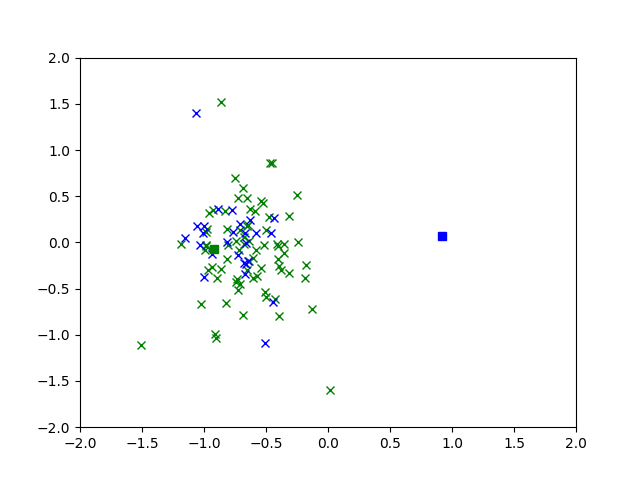}
    \end{subfigure}%
    \\
    \caption{[Best viewed in color] Evolution of system when the training process is biased towards $C_1$. Row, from top to bottom, represent time steps in the episode in increasing order. Columns, from left to right, correspond to the cases when both $C_1$ and $C_2$, only $C_1$, and only $C_2$ respectively, are allowed to broadcast messages. Members can always talk to their neighbors. Each cross represents a member's preference vector (which is two dimensional) and each square represents a candidate's propaganda vector (again, two dimensional). Members have been colored based on the candidate that they are following (green is for $C_1$). Note that the preference vector of all members tend to get close to propaganda vector of $C_1$. As preference vectors are updated only if a message is received, when one of the candidates is turned off, some preference vectors remain static.}
    \label{fig:biased_training_visualization}
\end{figure*}

\begin{figure*}
    \centering
    \begin{subfigure}[t]{0.33\textwidth}
        \centering
        \includegraphics[width=\linewidth]{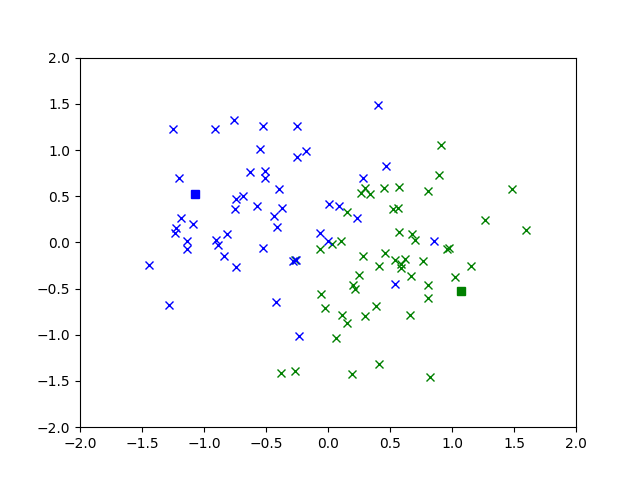}
    \end{subfigure}%
    ~ 
    \begin{subfigure}[t]{0.33\textwidth}
        \centering
        \includegraphics[width=\linewidth]{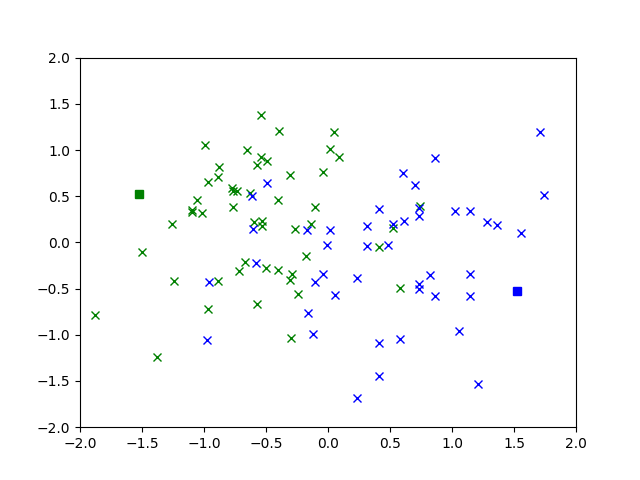}
    \end{subfigure}%
    ~
    \begin{subfigure}[t]{0.33\textwidth}
        \centering
        \includegraphics[width=\linewidth]{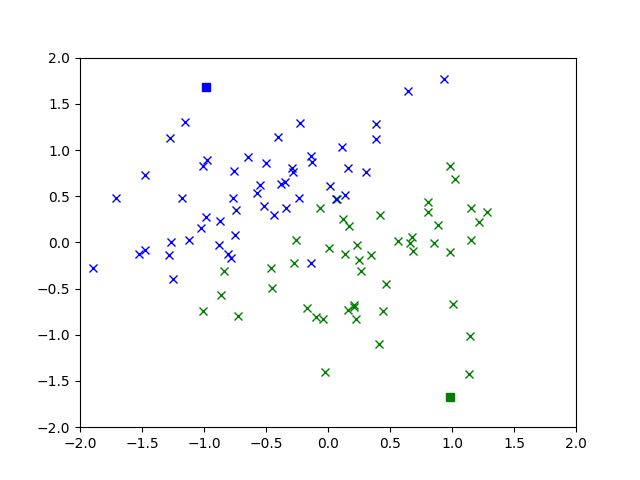}
    \end{subfigure}%
    \\
    \begin{subfigure}[t]{0.33\textwidth}
        \centering
        \includegraphics[width=\linewidth]{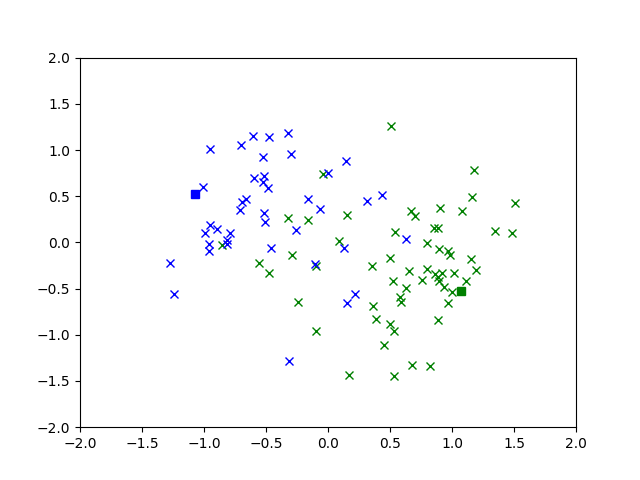}
    \end{subfigure}%
    ~ 
    \begin{subfigure}[t]{0.33\textwidth}
        \centering
        \includegraphics[width=\linewidth]{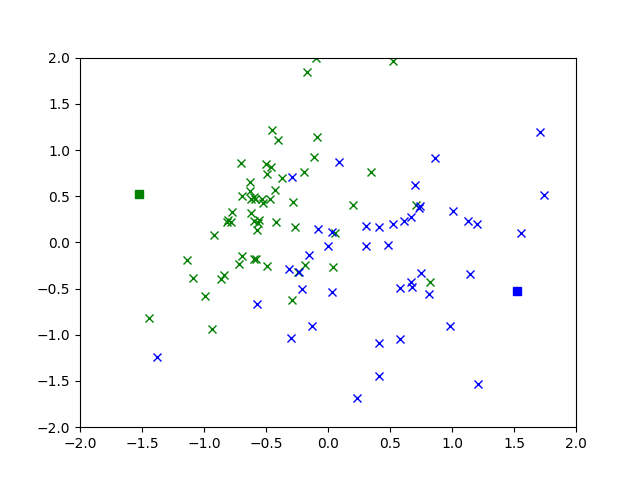}
    \end{subfigure}%
    ~
    \begin{subfigure}[t]{0.33\textwidth}
        \centering
        \includegraphics[width=\linewidth]{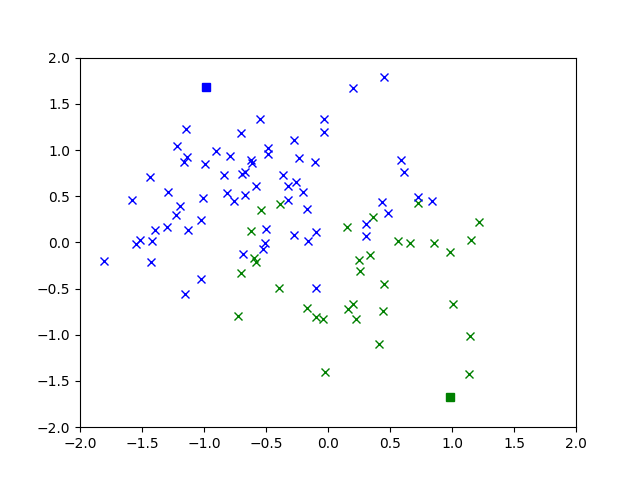}
    \end{subfigure}%
    \\
    \begin{subfigure}[t]{0.33\textwidth}
        \centering
        \includegraphics[width=\linewidth]{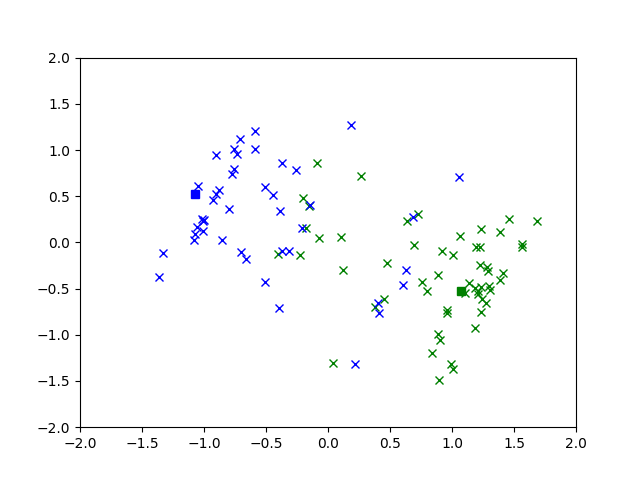}
    \end{subfigure}%
    ~ 
    \begin{subfigure}[t]{0.33\textwidth}
        \centering
        \includegraphics[width=\linewidth]{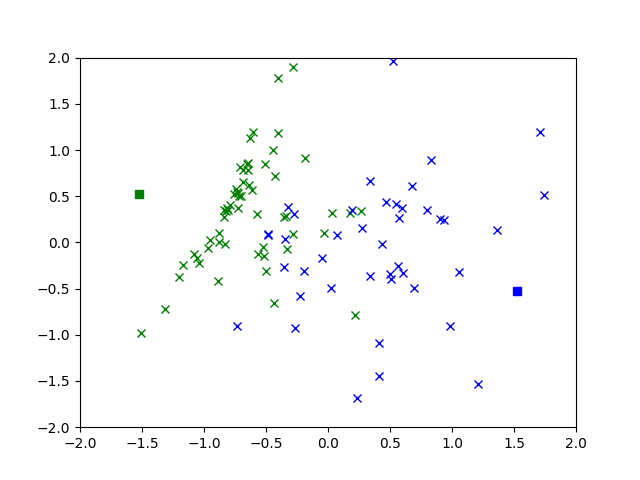}
    \end{subfigure}%
    ~
    \begin{subfigure}[t]{0.33\textwidth}
        \centering
        \includegraphics[width=\linewidth]{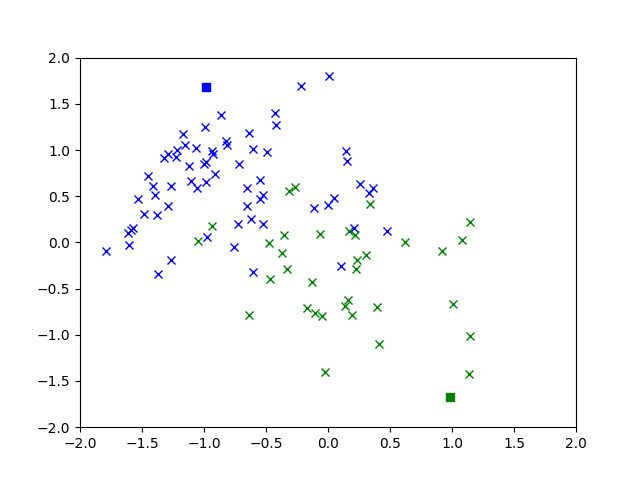}
    \end{subfigure}%
    \\
    \begin{subfigure}[t]{0.33\textwidth}
        \centering
        \includegraphics[width=\linewidth]{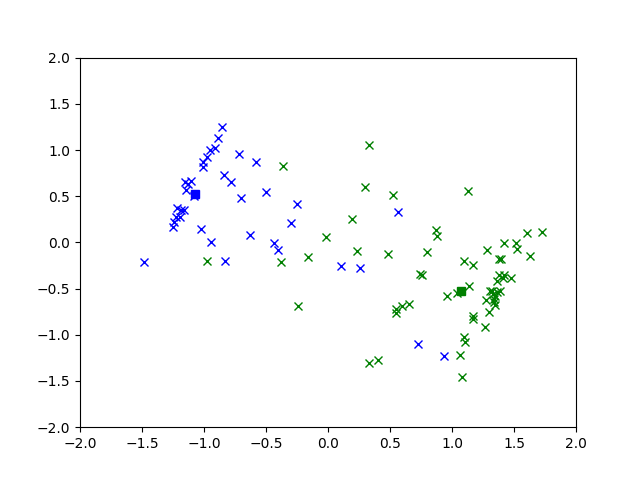}
    \end{subfigure}%
    ~ 
    \begin{subfigure}[t]{0.33\textwidth}
        \centering
        \includegraphics[width=\linewidth]{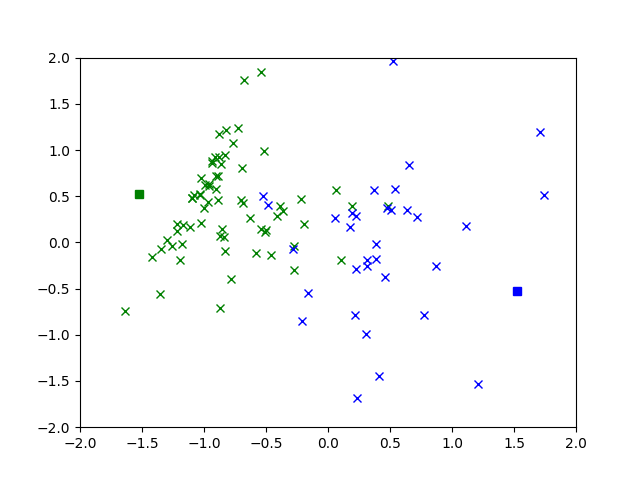}
    \end{subfigure}%
    ~
    \begin{subfigure}[t]{0.33\textwidth}
        \centering
        \includegraphics[width=\linewidth]{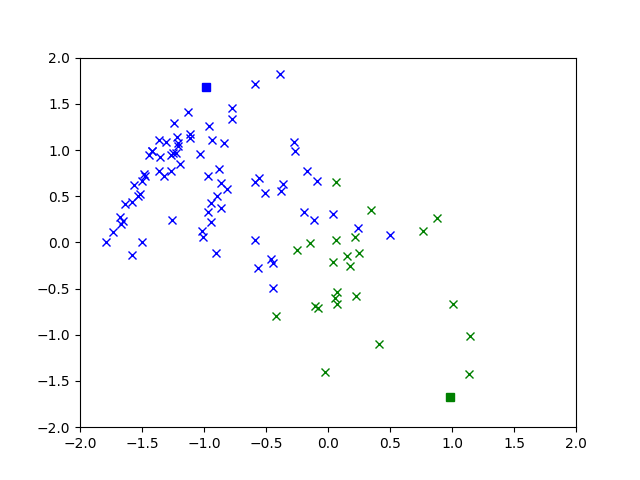}
    \end{subfigure}%
    \\
    \begin{subfigure}[t]{0.33\textwidth}
        \centering
        \includegraphics[width=\linewidth]{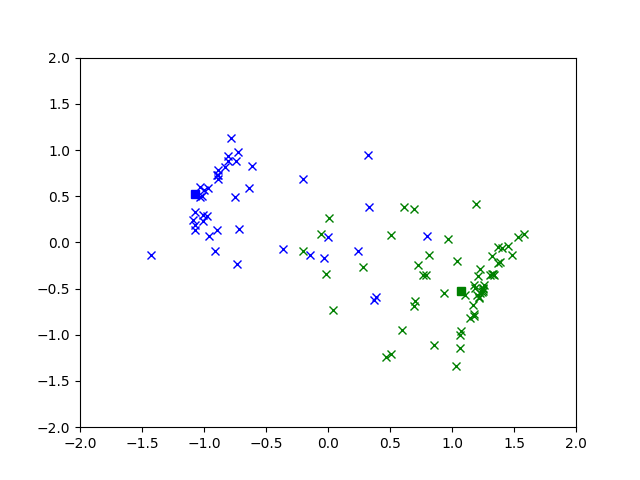}
    \end{subfigure}%
    ~ 
    \begin{subfigure}[t]{0.33\textwidth}
        \centering
        \includegraphics[width=\linewidth]{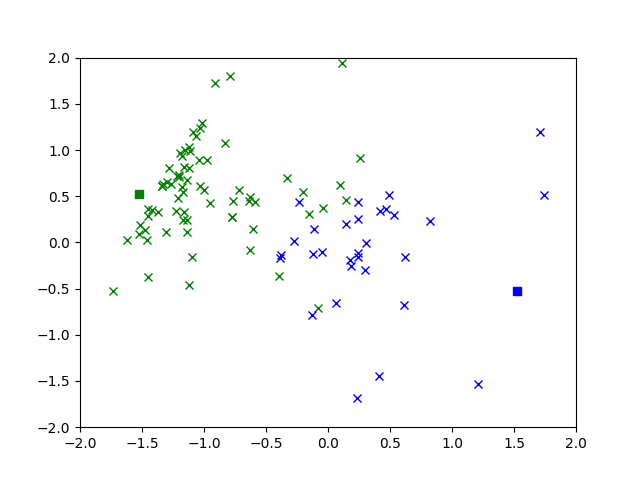}
    \end{subfigure}%
    ~
    \begin{subfigure}[t]{0.33\textwidth}
        \centering
        \includegraphics[width=\linewidth]{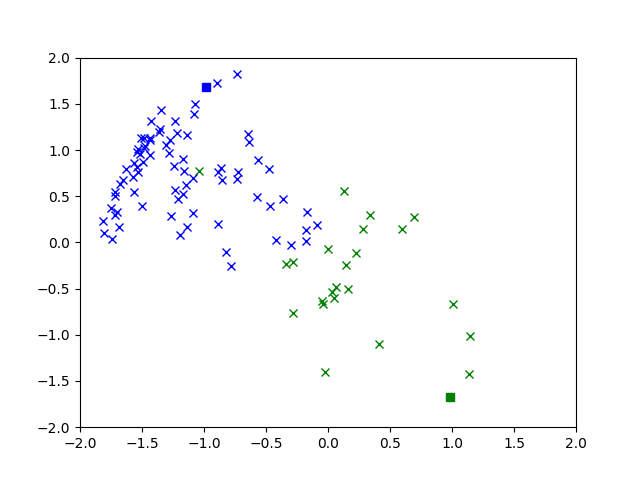}
    \end{subfigure}%
    \\
    \caption{[Best viewed in color] Evolution of preference vectors of members over time when training process is unbiased, i.e. each candidate is trying to win the game. The semantics are same as in Fig~\ref{fig:biased_training_visualization}.}
    \label{fig:unbiased_training_visualization}
\end{figure*}

In this section we visualize the evolution of preference vectors $\mathbf{m}_i^{(t)}$ over time. As noted earlier, the game consists of $T$ propaganda steps followed by a voting step. In Fig.~\ref{fig:biased_training_visualization}, we show how the preference vectors evolve during propaganda steps when the training process is biased towards candidate $C_1$ (both candidates are trying to make $C_1$ win). The case when training process is unbiased has been shown in Fig.~\ref{fig:unbiased_training_visualization}. Both the visualizations were obtained after training all the modules for $10000$ games using hyperparameters specified in Table~\ref{table:hyperparams}. Each game is played on a randomly sampled network of $100$ members as described in Section \ref{section:experiments}.

From Fig.~\ref{fig:biased_training_visualization}, it can be seen that preference vectors of members evolve over time to get close to propaganda vector of candidate $C_1$ as the training was biased in a way such that both candidates want $C_1$ to win. Additionally, when only one of the candidates is allowed to transmit, say $C_1$ to be more concrete, then the following observations can be made:
\begin{enumerate}
    \item Members that follow $C_2$ in the first time step do not move during a few initial time steps. This is because the preference vector is updated only if a message is received and when $C_2$ is not allowed to broadcast, these members do not receive any messages until one of their neighbors starts following $C_1$.
    \item Members communicate their intent to their neighbors as evident in the rightmost column where the members that were initially following $C_2$ eventually drift towards $C_1$. Since they do not get a direct message from $C_1$ and $C_2$ is not broadcasting, this can happen only if members communicate meaningfully with each other.
\end{enumerate}

In Fig.~\ref{fig:unbiased_training_visualization}, both candidates were trained to win the game and hence the training process is unbiased. It can be seen in the leftmost column that when both $C_1$ and $C_2$ are allowed to broadcast, then members partition themselves between $C_1$ and $C_2$. When only one of the candidates is active then the members change their preference vectors to get close to the propaganda vector of active candidate (middle and rightmost column). As before, by observing the rightmost column in Fig.~\ref{fig:unbiased_training_visualization}, it can be inferred that meaningful communication between nodes is taking place as when only $C_2$ is active, the members that were initially following $C_1$ eventually drift towards $C_2$ in response to the messages received from their neighbors.

\end{document}